\documentclass[final,3p,times]{elsarticle}
\usepackage{amssymb}
\usepackage{amssymb}
\usepackage[]{amsmath}
\usepackage{graphics}
\usepackage{amssymb}
\usepackage[]{amsmath}
\usepackage{amsthm}
\usepackage{setspace}
\usepackage{epsfig}
\usepackage{subfigure}
\usepackage{cases}
\usepackage{graphicx}

\biboptions{numbers,sort&compress}

\usepackage{amsmath}
\usepackage{times}
\usepackage{anysize}
\marginsize{2cm}{2cm}{0cm}{1.5cm}
\selectfont

\journal {}

\begin{document}

\begin{frontmatter}

\title{Modulation instability, rogue waves and spectral analysis for the sixth-order nonlinear Schr\"{o}dinger equation}

\author{ Yunfei Yue$^{a,b}$ }
\author{ Lili Huang$^{a,b}$ }
\author{ Yong Chen$^{a,b,c}$ \corref{cor1} }
\ead{ychen@sei.ecnu.edu.cn}

\cortext[cor1]{Corresponding author. }

\address[label1]{Institute of Computer Theory,  East China Normal University, Shanghai, 200062, China}

\address[label2]{Shanghai Key Laboratory of Trustworthy Computing, East China Normal University,
    Shanghai, 200062,  China}

\address[label3]{Department of Physics, Zhejiang Normal University,
    Jinhua, 321004,  China}

\begin{abstract}

Modulation instability, rogue wave and spectral analysis are investigated for the nonlinear Schr\"{o}dinger equation with the higher-order terms. The modulation instability distribution characteristics from the sixth-order to the eighth-order nonlinear Schr\"{o}dinger equations are studied. Higher-order dispersion terms are closely related to the distribution of modulation stability regime, and $n$-order dispersion term corresponds to $n-2$ modulation stability curves in the modulation instability band. Based on the generalized Darboux transformation method, the higher-order rational solutions are constructed. Then the compact algebraic expression of the $N$th-order rogue wave is given. Dynamic phenomena of first- to third-order rogue waves are illustrated, which exhibit meaningful structures. Two arbitrary parameters play important roles in the rogue wave solution. One can control deflection of crest of rogue wave and its width, while the other can cause the change of width and amplitude of rogue wave. When it comes to the third-order rogue wave, three typical nonlinear wave constructions, namely fundamental, circular and triangular are displayed and discussed. Through the spectral analysis on first-order rogue wave, when these parameters satisfy certain conditions, it occurs a transition between W-shaped soliton and rogue wave.

\end{abstract}

\end{frontmatter}

\section{Introduction}
It has been extensive interests in studying rogue waves in recent years. Rogue wave was first put forward conceptually in the ocean \cite{draperl-mo-1965}. Rogue waves are relatively large and spontaneous waves, whose appearance may result in catastrophic damage \cite{bwj-2006-kc-2009}. Large amplitude, unexpected, coming out from nowhere without warning and suddenly vanishing away without trace, are the basic characteristics \cite{akmhmedievn-pla-2009}. In general, the nonlinear partial differential equation satisfying the fiber communication model, for the stable solution, under the interaction of the dispersion term and the nonlinear term, there will be instability, which is called modulation instability (MI). MI is a basic nonlinear process of exponential increase with some small perturbations superimposed on the background of continuous waves in nonlinear dispersive media. MI of monochromatic nonlinear waves is a possible cause of the generation of rogue waves, which bursts sporadically more than the average level of the water surface. It is considered that MI is a most ubiquitous kind of instability in the nature world \cite{zve-pd-2009}. It exists both in the continuum and in the discrete nonlinear wave equations \cite{raptiz-pre-2004}.
Since Benjamin and Feir's groundbreaking hydrodynamics construction \cite{bjf-1967}, MI has played a prominent role in diverse areas of scientific research, for example, plasma physics \cite{taniutit-prl-1968}, nonlinear optics \cite{bespalovvi-1966}, and fluid dynamics \cite{withamgb-1965}.

In fact, the above mentioned instability can result in self-induced modulation of incoming continuous waves with subsequent local pulses, which may be discovered in many physical systems. Due to the presence of this phenomenon, there are many interesting physical effects, such as break-up of deep water-gravity waves in the ocean, the formation of envelope solitons in electrical transmission lines and optical fibers, as well as the formation of cavitons in plasmas. Different distributions of the MI gain can lead to distinct pattern of nonlinear dynamic phenomena \cite{zhaolc-pre-2014}. The dispersion term and the nonlinear term are playing different roles in the nonlinear systems, but both of them affect the instability of the solutions for the nonlinear systems. Recently, some literatures have analyzed the importance of high-order dispersion terms, which is not only affect MI \cite{jhzhang-prsa-2017} but also induce some novel excited states \cite{wangx-jmaa-2017,wangl-pre-2016}. The study of MI regions in nonlinear systems is crucial in many fields and is the basis for interpreting or regulating various models or phenomena in different fields.

The nonlinear Schr\"{o}dinger (NLS henceforth) equation has a prominent position in nonlinear physics. It has extensive physical applications, especially in nonlinear optics \cite{gpagrawal-nfo-2006}, atmosphere \cite{stenflol-jpp-2010}, and  water waves \cite{aosborne-2010}. In 1983, Peregrine \cite{peregrinedh-jams-1983} gave the analytical expression of the rogue wave in the first-order as an outcome of MI on the constant wave background. This type of rogue wave also has another name, that is Peregrine breather. In recent years, many authors \cite{akhmediewn-pre-2009,Ankiewicza-jpagp-2010,ohtay-prsampes-2012} have reported the higher-order rogue wave solutions, some important physical properties and applications for the NLS equation. In addition, various extensions of NLS equation have also been studied, such as pair-transition-coupled NLS equation \cite{zhanggq-prsa-2017}, variable coefficient NLS equation  \cite{gagnonl-jpa-1993,wangl-pre-2016-1,yangyq-nd-2015}, three-component NLS equations\cite{zhanggq-cnsns-2018}, three-component coupled derivative NLS equations \cite{xut-nd-2017}, and $n$-component NLS equations \cite{zhanggq-prsa-2018}. General high-order solitons of three different types of nonlocal NLS equations in the reverse-time, PT-symmetric and reverse-space-time were derived by using a Riemann-Hilbert treatment \cite{yangb-nd-2018}.

However, there exist only lowest-order terms (dispersion and nonlinearity) in the standard NLS equation \cite{ankiewicza-pre-2014}. When the characteristics of the solutions exceed the
simple approximation in deriving the NLS equation, the higher-order terms will hold the dominate role \cite{caily-nd-2017}. For instance, it may help to illustrate the physical properties of wave blow-up and collapse \cite{berge-pr-1998}. In 2016, Ankiewicz \cite{ankiewicza-pre-2016} \emph{et al}. studied the following form of NLS equation containing higher-order nonlinear terms and dispersion terms,

\begin{equation}\label{eq-1.1}
iq_z+\delta_2\Gamma_2(q)-i\delta_3\Gamma_3(q)+\delta_4\Gamma_4(q)-i\delta_5\Gamma_5(q)+\delta_6\Gamma_6(q)-i\delta_7\Gamma_7(q)+\delta_8\Gamma_8(q)+\cdots=0,
\end{equation}
with
\begin{equation*}
\begin{split}
 \Gamma_2 =& q_{tt}+2q|q|^2, \\
  \Gamma_3 =& q_{ttt}+6{q}^2q_t,  \\
  \Gamma_4 =& q_{tttt}+6q^{*}q_t^2+4q|q_t|^2+8|q|^2q_{tt}+2q^2q_{tt}^{*}+6|q|^4q, \\
  \Gamma_5 =& q_{ttttt}+10|q|^2q_{ttt}+30|q|^4q_t+10qq_tq_{tt}^{*}+10qq_t^{*}q_{tt}+20q^{*}q_tq_{tt}+10q_t^2q_t^{*},\\
\Gamma_6 =& q_{tttttt}+q^2[60|q_t|^2q^{*}+50q_{tt}(q^{*})^2+2q_{tttt}^{*}]+q[12q^{*}q_{tttt}+18q^{*}_tq_{ttt}+8q_tq^{*}_{ttt}+70(q^{*})^2q_t^2+22|q_{tt}|^2]\\
&+10q_t[3q^{*}q_{ttt}+5q^{*}q_{tt}+20q^{*}q_{tt}^2]+10q^3[2q^{*}q_{tt}^{*}+(q_t^{*})^2]+20q^{*}q_{tt}^2+20q|q|^6,\\
\Gamma_7 =& q_{ttttttt}+70q_{tt}^2q_t^{*}+112q_t|q_{tt}|^2+98|q_t|^2q_{ttt}+70q^2\{q_t[2q^{*}q_{tt}^{*}+(q_t^{*})^2]+q^{*}(2q_{tt}q_t^{*}+q_{ttt}q^{*})\}\\
& +28q_t^2q_{ttt}^{*}+14q[q^{*}(20|q_t|^2q_t+q_{ttttt})+3q_{ttt}q_{tt}^{*}+2q_{tt}q_{ttt}^{*}+2q_{tttt}q_t^{*}+q_t+q_{tttt}^{*}+20q_tq_{tt}(q^{*})^2]\\
& +140|q|^6q_t+70q_t^3(q^{*})^2+14(5q_{tt}q_{ttt}
+3q_tq_{tttt})q^{*},\\
\Gamma_8 = & q_{tttttttt}+14q^3[40|q_t|^2(q^{*})^2+20(q^{*})^3q_{tt}+2q^{*}q_{tttt}^{*}+4q_t^{*}q_{ttt}^{*}+3(q_{tt}^{*})^2]\\
& +q^2[28q^{*}(14|q_{tt}|^2+6q_tq_{ttt}^{*}+11q_t^{*}q_{ttt}+238q_{tt}(q_t^{*})^2+336|q_t|^2q_{tt}^{*}+560q_t^2(q^{*})^3\\
& +98q_{tttt}(q^{*})^2+2q_{tttttt}^{*}]+2q\{21q_t^2[9(q^{*})^2+14q^{*}q_{tt}^{*}]+q_t[728q_{tt}q_t^{*}q^{*}+238q_{ttt}(q^{*})^2\\
& +6q_{ttttt}^{*}]+34|q_{ttt}|^2+36q_{tttt}q_{tt}^{*}+22q_{tt}q_{tttt}^{*}+20q_{ttttt}q_t^{*}+161q_{tt}^2(q^{*})^2+8q_{tttttt}q^{*}\}\\
 & +182q_{tt}|q_{tt}|^2+308q_{tt}q_{ttt}q_t^{*}+252q_tq_{ttt}q_{tt}^{*}+196q_tq_{tt}q_{ttt}^{*}+168q_tq_{tttt}q_t^{*}+42q_t^2q_{tttt}^{*}\\
 & +14q^{*}(30q_t^3q_t^{*}+4q_{ttttt}q_t+5q_{ttt}^2+8q_{tt}q_{tttt})+490(q^{*})^2q_t^2q_{tt}+140q^4q^{*}[q^{*}q_{tt}^{*}+(q_t^{*})^2]+70q|q|^8,\\
\end{split}
\end{equation*}
where $|q|=|q(z,t)|$ denotes envelope of the optical pulse with spatial coordinate $z$ and scaled time coordinate $t$, $\delta_i$ ($i=2,3,4,5,\cdots\infty$) represents the $i$-order real dispersion coefficient. $\Gamma_3$ is the Hirota operator \cite{hirotar-jmp-1973}, $\Gamma_4$ is the Lakshmanan-Porsezian-Daniel operator \cite{lakshmananm-pla-1988}, $\Gamma_5$ is known as the quintic operator \cite{chowdurya-pre-2014}, $\Gamma_6$ is the sextic operator, $\Gamma_7$ is the heptic operator, $\Gamma_8$ is the octic operator. With an infinite number of arbitrary coefficients, these extensions are integrable. The arbitrariness of coefficients enables us to go well beyond the single NLS equation.

One of our goals in this paper is to investigate MI of a continuous wave for the NLS equation (\ref{eq-1.1}) with different higher-order terms. We discuss MI distribution characteristics from the sixth-order NLS equation to the eighth-order NLS equation. Comparing their MI gain functions of NLS equations with different order dispersion terms, it enables us to find the distribution law of MS curves in the MI band. Then we focus on study rogue waves of the following reduced sixth-order NLS equation \cite{sunjj-epjp-2017,sunwr-adp-2017,lanzz-oqe-2018}
\begin{equation}\label{eq-1.2}
iq_z+\delta_2\Gamma_2(q)+\delta_6\Gamma_6(q)=0.
\end{equation}
Nowadays many methods have been developed to investigate rogue waves of the nonlinear systems, such as the Darboux transformation (DT) \cite{gbl-pre-2012,hejs-pre-2013,wenxy-chaos-2015,wenxy-jmp-2018,weij-cnsns-2018,linglm-pd-2016}, Hirota method \cite{liuyb-nd-2017,chenjc-jmaa-2018,huangll-cma-2018,yueyf-aml-2019}, nonlocal symmetry method \cite{huangll-cnsns-2019}.
Based on the generalized DT, higher-order rogue waves will be generated for Eq. (\ref{eq-1.2}). For the parameter of $\delta_2$, many papers \cite{akhmedievn-1997,ankiewicza-pre-2010,ankiewicza-pla-2014} choose $\delta_2=\frac{1}{2}$, this setting has certain convenient features.
Here, we also set $\delta_2=\frac{1}{2}$ in Eq. (\ref{eq-1.2}). Via the analytical rational expressions and MI characteristics, the dynamics of rogue waves will be studied in detail.

It is also our purpose here to investigate how to use the spectral features of the propagating wave envelope to reveal the existence of nonlinearity and rogue wave in a short time before the occurrence of a special rogue wave event. To establish the results, we apply the spectral analysis approach  \cite{akhmediev-pla-2011,akhmediev-pd-2015,bayindir-pre-2016,wangx-chaos-2017,wangx-sam-2017} to the first-order rogue wave solutions of Eq. (\ref{eq-1.2}).

The remainder of our article is constructed as follows. In Section 2, MI distribution features of the NLS equation with different higher-order  nonlinear and dispersion terms will be discussed according to MI analysis theory. By virtue of the generalized DT of the sixth-order NLS equation, a concrete expression of the N-order rogue wave solutions will be given in Section 3.  In Section 4, Utilizing the expressions obtained in the previous section, the first-order, second-order, and third-order exact rogue wave solutions are presented, where their dynamic behavior are also analyzed. Section 5 is devoted to spectral analysis on the first-order rogue wave. Finally, some conclusions are given.


\section{Modulation instability}

MI is observed in a time-averaged way and usually triggered from a continuous wave or quasi-continuous wave. The continuous wave condition is corresponding to an effectively unbounded MI domain. Then it can yield information on average behavior of the nonlinear process and the general tendencies for instability, but usually prevents time-resolved of the stochastic dynamics. MI symmetry breaking can occur for the reason of higher-order dispersion \cite{droquesm-ol-2011}. MI is the basic mechanism for generating of rogue wave solutions. MI is an interactive gain procedure that generates priority frequency intervals between patterns \cite{sollidr-np-2012}.
Studied here is the MI analysis on continuous waves for the NLS equation with different higher-order dispersion terms, in order to reveal the MI features considering of higher-order dispersion effects. The plane wave solution of system (\ref{eq-1.1}) has the following form
\begin{equation}\label{eq-4.1}
  q_{cw}=Ae^{i\theta}=Ae^{i(kz+\omega t)}.
\end{equation}
There are three real constants, wave number $k$, amplitude $A$ and frequency of background $\omega$. Substituting Eq. (\ref{eq-4.1}) into Eq. (\ref{eq-1.2}),  it can be obtained that
\begin{equation}\label{eq-4.1k}
k=20A^6\delta_6-90A^4\omega^2\delta_6+30A^2\omega^4\delta_6-\omega^6\delta_6+A^2-\frac{1}{2}\omega^2.
\end{equation}
According to the MI theory, we add a small perturbation function $p(z,t)$ to the plane-wave solution. Then a perturbation solution can be derived as
\begin{equation}\label{eq-4.2}
  q_{pert}=\left(A+\epsilon p(z,t)\right)e^{i(kz+\omega t)},
\end{equation}
where $p(z,t)=me^{i(Kz+\Omega t)}+ne^{-i(Kz+\Omega t)}$, $\Omega$ indicates the disturbance frequency, and $m,n$ are both small parameters. Substituting the perturbation solution (\ref{eq-4.2}) into the sixth-order NLS equation (\ref{eq-1.2}), it can generate a system of linear homogeneous equations for $m$ and $n$. Based on the existence conditions for solutions of linear homogeneous equations, that is, the determinant of coefficient matrix of $m$ and $n$ is equal to 0, which gives rise to a dispersion relation equation. By solving this dispersion relation equation, MI gain can be obtained
\begin{equation}\label{eq-4.3}
\begin{split}
  G_6 &= |Im(K)|= \frac{1}{2}Im\left(\Omega\sqrt{(\Omega^2-4A^2)g_6^2}\right),\\
  g_6 &= 1+\left[2\Omega^4+(-20A^2+30\omega^2)\Omega^2+60A^4-180A^2\omega^2+30\omega^4\right]\delta_6.
\end{split}
\end{equation}

Similar to the above calculation process, we also obtain the MI gain functions of the seventh-order (i.e. $\delta_2=\frac{1}{2},\delta_7\neq0,\delta_3\cdots\delta_6=0$ in Eq. (\ref{eq-1.1})) and eighth-order (i.e. $\delta_2=\frac{1}{2},\delta_8\neq0,\delta_3\cdots\delta_7=0$ in Eq. (\ref{eq-1.1})) NLS equation, respectively. Their exact expressions are as follows

\begin{equation}\label{eq-4.4}
\begin{split}
  G_7 &= |Im(K)|= \frac{1}{2}Im\left(\Omega\sqrt{(\Omega^2-4A^2)g_7^2}\right),\\
  g_7 &= 1+\left[14\,\omega\,{\Omega}^{4}+ \left( -140\,{A}^{2}\omega+70\,
{\omega}^{3} \right) {\Omega}^{2}+420\,{A}^{4}\omega-420\,{A}^{2}{
\omega}^{3}+42\,{\omega}^{5}\right]\delta_7,
\end{split}
\end{equation}
and
\begin{equation}\label{eq-4.5}
\begin{split}
  G_8 =& |Im(K)|  = \frac{1}{2}Im\left(\Omega\sqrt{\left(\Omega^2-4A^2\right)g_8^2}\right),\\
  g_8 =& 1+\Big[-2\,{\Omega}^{6}+\left( 28\,{A}^{2}-56\,{\omega}^{2} \right) {\Omega}^{4}+ \left( -140\,{A}^{4}+560\,{A}^{2}{\omega}^{2}-
140\,{\omega}^{4}\right) {\Omega}^{2}\\
    & ~~+280\,{A}^{6}-1680\,{A}^{4}{\omega}^{2}+840\,{A}^{2}{\omega}^{4}-56\,{\omega}^{6}\Big]\delta_8.
\end{split}
\end{equation}

From the above analysis, it appears that there exists two distinctive MI and modulation stability (MS) regions. In the region of $|\Omega|<2A$, MI exists when $g_i\neq0, (i=6,7,8)$. On the contrary, if $g_i=0, (i=6,7,8)$, there appears nontrivial features in the MI region. This in turn implies that a MS region occurs in the region of low perturbation frequency, where the growth rate of corresponding MI decays to zero.

\begin{figure*}[!htbp]
\centering
\subfigure[]{\includegraphics[height=1.5in,width=1.6in]{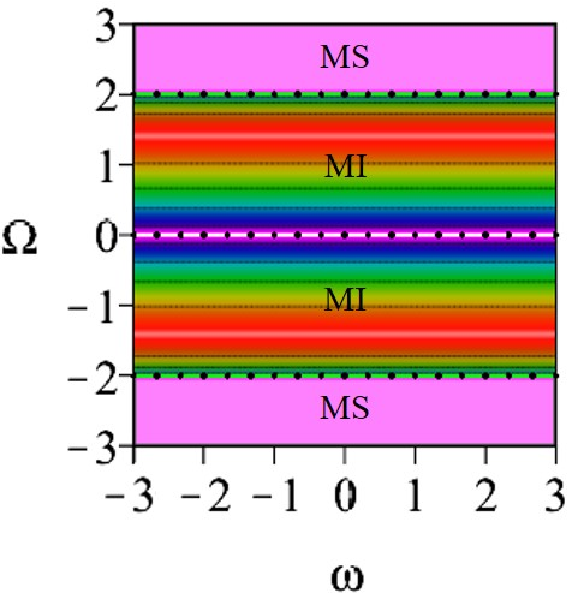}}
\subfigure[]{\includegraphics[height=1.5in,width=1.6in]{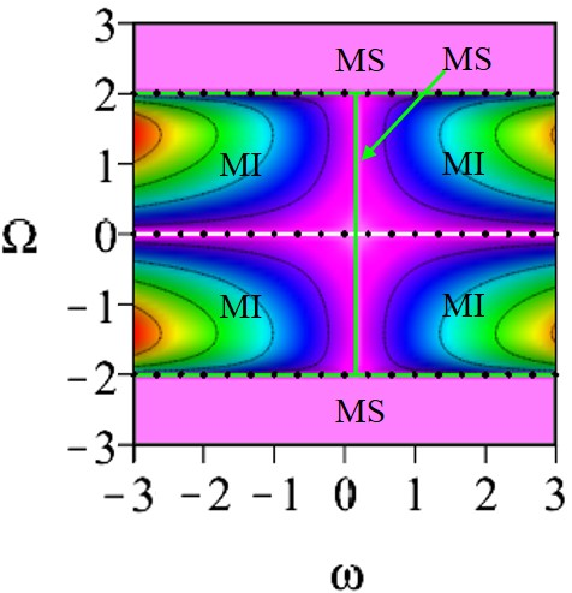}}
\subfigure[]{\includegraphics[height=1.5in,width=1.6in]{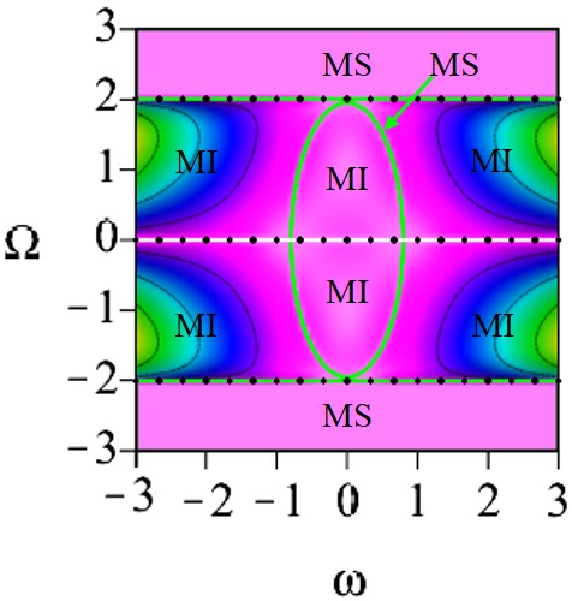}}
\subfigure[]{\includegraphics[height=1.5in,width=1.6in]{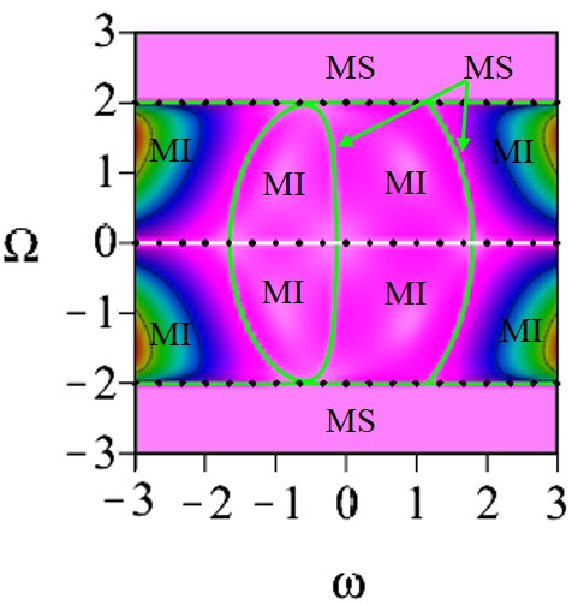}}
\subfigure[]{\includegraphics[height=1.5in,width=1.6in]{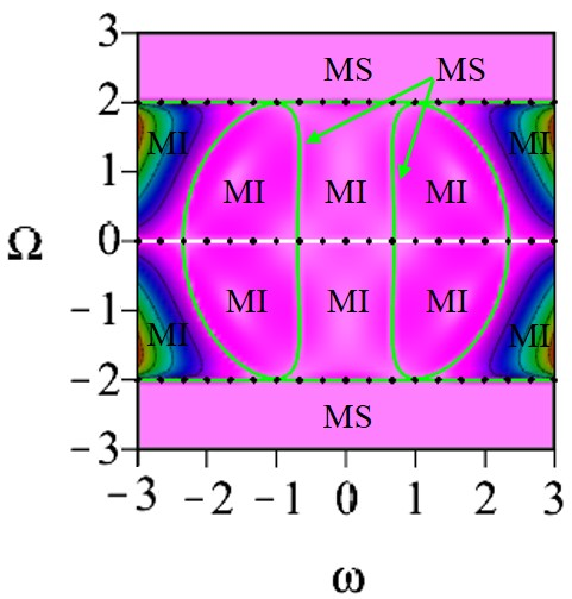}}
\subfigure[]{\includegraphics[height=1.5in,width=1.6in]{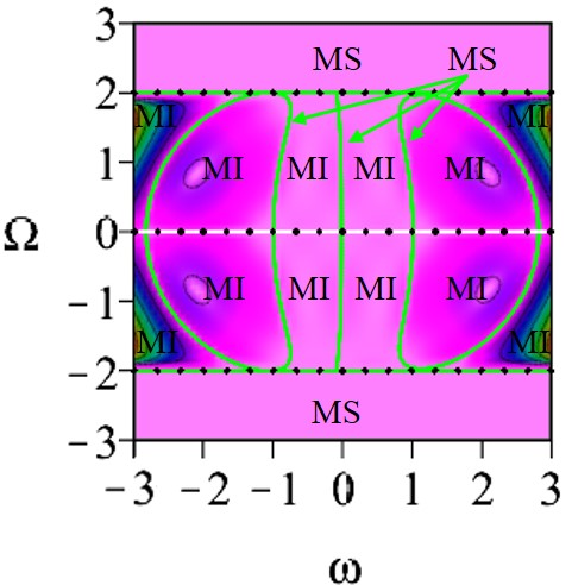}}
\subfigure[]{\includegraphics[height=1.5in,width=1.6in]{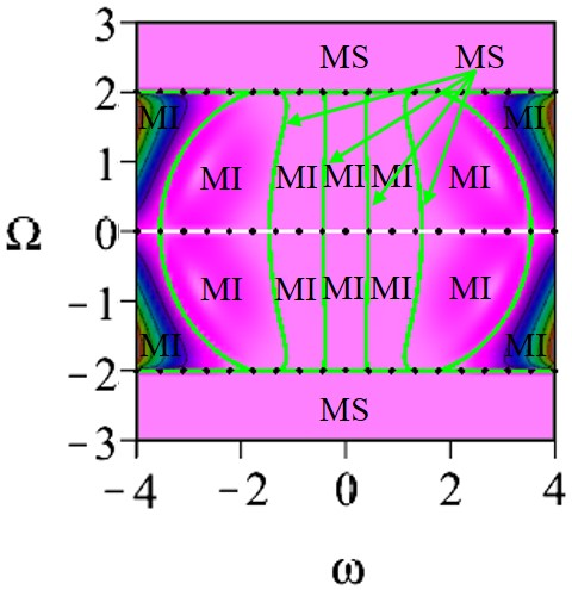}}
\caption{ Plots of the distribution of MI gain with perturbation frequency $\Omega$ and continuous background frequency $\omega$, and $A=1$. The dashed white lines indicate the resonance lines, the dashed green lines mean boundary lines, and the solid green lines represent that perturbation is stable. In addition to MS curves and MS quasi-elliptic curves, the remaining areas are all non-zero MI gain in MI band. (a) The standard NLS equation \cite{zhaolc-josa-2016}: no MS region exists in the MI band. (b) The Hirota equation \cite{liuc-pre-2016}: an MS curve exists in the MI band. (c) The Lakshmanan-Porsezian-Daniel equation \cite{duanl-pre-2017} with $\delta_3=0$: an MS elliptic ring appears in the MI band. (d) The fifth-order NLS equation \cite{yangyq-chaos-2015,lip-aml-2018} with $\delta_3=\delta_4=0$: it not only has an  MS curve, but also has an MI quasi-elliptic ring in MI band. (e) The sixth-order NLS equation: it has two MS quasi-elliptic rings in the MI band. (f) The seventh-order NLS equation: here exists an MS curve and two MS quasi-elliptic rings in the MI band.  (g) The eighth-order NLS equation: two MS curves and two MS quasi-elliptic rings appear in the MI band. }
\label{Fig-4.1}
\end{figure*}

Comparing their MI gain functions of NLS equations with different order dispersion terms, we can obtain the distribution law of MS curves in the MI band.

$\bullet$~~When $\delta_2=\frac{1}{2}$ and the remaining $\delta_i=0,i=3,4,5,\cdots$, the system (\ref{eq-1.1}) is reduced to classical NLS equation. And the highest power of $g_2(\omega)$ is equal to 0, namely, $g_2(\omega)=1$. Therefore, no MS region exists in the MI band ($|\Omega|<2A$), which is described in Fig. \ref{Fig-4.1}(a).

$\bullet$~~When the above-described conditions wherein $\delta_3=0$ becomes $\delta_3\neq0$, the system  (\ref{eq-1.1}) is transformed into third-order NLS equation. And the highest power of $g_3(\omega)$ is 1, i.e. a simple factor of $\omega$. It appears that an MS curve exists in the MI band ($|\Omega|<2A$), which is described in Fig. \ref{Fig-4.1}(b).

$\bullet$~~When the above-described conditions wherein $\delta_4=0$ becomes $\delta_4\neq0$, the system (\ref{eq-1.1}) can be degenerated to fourth-order NLS equation. The highest power of $g_4(\omega)$ is 2. There exists an MS elliptic ring in the MI band ($|\Omega|<2A$), which is described in Fig. \ref{Fig-4.1}(c).

$\bullet$~~When the above-described conditions wherein $\delta_5=0$ becomes $\delta_5\neq0$, then we can transform (\ref{eq-1.1}) into fifth-order NLS equation. The highest power of $g_5(\omega)$ is 3. Both an MS curve and an MS quasi-elliptic ring occur in the MI band ($|\Omega|<2A$), which is illustrated in Fig. \ref{Fig-4.1}(d).

$\bullet$~~When the above-described conditions wherein $\delta_6=0$ becomes $\delta_6\neq0$, we can transform (\ref{eq-1.1}) into sixth-order NLS equation. And the highest power of $g_6(\omega)$ is 4. There are two MS quasi-elliptic rings in the MI band ($|\Omega|<2A$), which is illustrated in Fig. \ref{Fig-4.1}(e).

$\bullet$~~When the above-described conditions wherein $\delta_7=0$ becomes $\delta_7\neq0$, and Eq. (\ref{eq-1.1}) is reduced to seventh-order NLS equation. The highest power of $g_7(\omega)$ is 5. There exists an MS curve and two MS quasi-elliptic rings in the MI band. ($|\Omega|<2A$), see Fig. \ref{Fig-4.1}(f).

$\bullet$~~When the above-described conditions wherein $\delta_8=0$ becomes $\delta_8\neq0$, we can transform (\ref{eq-1.1}) into eighth-order NLS equation. And the highest power of $g_8(\omega)$ is 6. Then it reveals that two MS curves and two MS quasi-elliptic rings exist in the MI band ($|\Omega|<2A$). The distribution of this case is illustrated in Fig. \ref{Fig-4.1}(g).

The MI distribution features of all above higher-order dispersion NLS equations are illustrated by Figs. \ref{Fig-4.1}(a-g). According to the above analysis process, it is evident that there exist two arbitrary parameters, namely higher order dispersion coefficient $\delta_i,i=2,3,4\cdots$ and amplitude $A$. These parameters control the MS distribution of system (\ref{eq-1.1}) in MI band. By adjusting the parameters, the MS quasi-elliptic and MS elliptic ring can be completely contained within the MI band or intersected at the MI boundary, the latter case yields two curves in MI band.
By analyzing the expressions in Eq. (\ref{eq-4.3}), we can discuss the MI distribution characteristics of the sixth-order NLS equation (\ref{eq-1.2}). Obviously, $g_6$ is a polynomial about $\omega$ and its highest power is 4. If this polynomial factor is decomposed into the product form of a single factor, then we can get four solutions, which shows that $G_6$ has four curves in the frequency plane $(\omega,\Omega)$. Here, Fig. \ref{Fig-4.1}(e) illustrates the MI gain distribution features in the frequency plane $(\omega,\Omega)$. It is clear that this frequency plane contains two different regions, namely, MI and MS. The expression $\Omega^2-4A^2$ in $G_6$ indicates that a low-perturbed frequency MI band ($|\Omega|<2A$) exists in the frequency plane $(\omega,\Omega)$. Setting $g_6=0$, it gives rise to MI gain $G=0$, which represents two MS quasi-elliptic rings in frequency plane $(\omega,\Omega)$ and demonstrated by Fig. \ref{Fig-4.1}(e). When selecting suitable parameters so that both elliptical semi-major axis is greater than 2, then there exist four curves in the MI band. With selecting of $\delta_6=0$, Eq. (\ref{eq-1.2}) is then transformed into classical NLS equation and the corresponding MI gain $G_6$ is reduced to $ \frac{1}{2}Im\left(\Omega\sqrt{(\Omega^2-4A^2)}\right)$. Fig. \ref{Fig-4.1}(a) displays the MI gain distribution features of classical NLS equation. Quite evidently, there are neither MS curves  nor MS quasi-elliptic rings in the MI band.

It is easy to find that Fig. \ref{Fig-4.1} only shows one case where the MS ellipse is contained in the MI band, i.e. the semi-long axis of the MS ellipse less than 2. When choosing appropriate values of parameters to make the semi-long axis of the MS ellipse greater than 2, we can get another case that only the MS curves exists in the MI band. From Fig. \ref{Fig-4.1}(g), there appears that the MS elliptic ring in the middle degenerates into two MS curves. From the standard NLS equation to eighth-order dispersion NLS equation, the number of the MS curves are $0,1,2,3,4,5,6,$ respectively; and the highest power of $g_i(\omega),(i=2,3,\ldots)$ are $0,1,2,3,4,5,6,$ respectively. So we can obtain relation among them, which is listed in Table \ref{Tab-1}.

\begin{table}
\caption{Relation between three parameters}
\centering
\begin{tabular}{|c|c|c|c|c|c|c|c|c|c|}
\hline
\multicolumn{10}{|c|}{The Nonlinear Schr\"{o}dingner equation with higher-order terms}              \\ \hline
The order of dispersion term               & 2 & 3 & 4 & 5 & 6 & 7 & 8 & ... & n   \\ \hline
The highest power of $\omega$ in $g_i$ function & 0 & 1 & 2 & 3 & 4 & 5 & 6 & ... & n-2 \\ \hline
The number of the MS curves in MI band     & 0 & 1 & 2 & 3 & 4 & 5 & 6 & ... & n-2 \\ \hline
\end{tabular}
\label{Tab-1}
\end{table}

\section{Gneralized Darboux Transformation for the sixth-order NLS equation}

So as to study rational solution for the sixth-order NLS equation (\ref{eq-1.2}), we will construct a generalized DT in this section. Starting from the following linear spectral problems of Eq. (\ref{eq-1.2}), namely,

\begin{equation}\label{eq-2.1}
\begin{split}
\Psi_t &= {i(\lambda\sigma_1+Q)}\Psi,  \\
\Psi_z &= \sum_{c=0}^6i\lambda^cV_{c},\\
\end{split}
\end{equation}
where
\begin{equation*}
  \sigma_1=
  \begin{pmatrix}
    1 & 0 \\
    0 & 1
    \end{pmatrix},~~~~
    Q=\begin{pmatrix}
    0 & q^* \\
    q & 0
    \end{pmatrix},~~~~
    V_c=\begin{pmatrix}
    A_{c} & B_{c}^{*}\\
    B_{c} & -A_{c}
    \end{pmatrix},\\
\end{equation*}
with
\begin{equation*}
  \begin{split}
  A_{0} &= -\frac{1}{2}|q|^2-10\delta_6|q|^6-5\delta_6[q^2(q_t^*)^2+(q^*)^2q_t^2]-10\delta_6|q|^2(qq_{tt}^*+q^*q_{tt})
  -\delta_6|q_{tt}|^2+\delta_6(q_tq_{ttt}^*+q_t^*q_{ttt}-q^*q_{tttt}
  -qq_{tttt}^*),  \\
  A_{1} &= 12i\delta_6|q|^2(q_tq^*-q_t^*q)+2i\delta_6(q_tq_{tt}^*
  -q_t^*q_{tt}+q^*q_{ttt}-q_{ttt}^*q),~~A_{2} = 1+12\delta_6|q|^4-4\delta_6|q_t|^2+4\delta_6(q_{tt}^*q+q_{tt}q^*),\\
  A_{3} &= 8i\delta_6(qq_t^*-q^*q_t),~~A_{4} = -16\delta_6|q|^2,~~A_{5} = 0,~~A_{6} = 32\delta_6,~~B_{2} = -24i\delta_6|q|^2q_t-4i\delta_6q_{ttt},\\
  B_{0} &= \frac{i}{2}q_t+i\delta_6q_{ttttt}+10i\delta_6(qq_t^*q_{tt}+qq_{tt}^*q_t+|q|^2q_{ttt}
  +3|q|^4q_t+q_t|q_t|^2+2q^*q_tq_{tt}),\\
  B_{1} &= q+12\delta_6q^*q_t^2+16\delta_6|q|^2q_{tt}+4\delta_6q^2q_{tt}^*+2\delta_6q_{tttt}+12\delta|q|^4q+8\delta_6q|q_t|^2,\\
  B_{3} &= -16\delta_6|q|^2q-8\delta_6q_{tt}, ~~B_{4} = 16i\delta_6q_t,~~B_{5} = 32\delta_6q,~~B_{6} = 0,
  \end{split}
\end{equation*}
and eigenfunction $\Psi=(\psi_1,\phi_1)^{\dag}$, $\psi_1$ and $\phi_1$ denote complex functions with $z$ and $t$, $\dag$ means matrix transpose, $\ast$ denotes the complex conjugation, $\lambda$ is an spectral parameter of the linear spectral problem (\ref{eq-2.1}). It is clearly that $U_z-V_t+UV-VU=0$, which is the compatibility condition of (\ref{eq-2.1}), can directly give rise to Eq. (\ref{eq-1.2}).

Supposing that $\Psi=(\psi_1,\phi_1)\dag$ is a fundamental solution based on the above spectral problem. Then the basic DT for Eq. (\ref{eq-1.2}) has the form

\begin{equation}\label{eq-2.2}
\begin{split}
  \Psi[1] &= T[1]\Psi,~~T[1] = \lambda I-H[0]\Lambda_1H[0]^{-1},\\
  q[1] &= q[0]+2(\lambda_1^{*}-\lambda_1)\frac{\psi_1[0]^{*}\phi_1[0]}{(|\psi_1[0]|^2+|\phi_1[0]|^2)},
\end{split}
\end{equation}
where $\phi_1[0]=\phi_1,~~\psi_1[0]=\psi_1$, and
\begin{equation*}
\begin{gathered}
  I=
  \begin{pmatrix}
   1 & 0  \\
    0 & 1
    \end{pmatrix},~~
  H[0]=
  \begin{pmatrix}
   \psi_1[0] & -\phi_1[0]^{*}  \\
    \phi_1[0] & \psi_1[0]^{*}
    \end{pmatrix},~~
  \Lambda_1=
  \begin{pmatrix}
   \lambda_1 & 0  \\
    0 & \lambda_1^{*}
    \end{pmatrix}.
\end{gathered}
\end{equation*}

In the general case, assuming similarly that $\Psi_l=(\psi_l,\phi_l)^{\dag}$, $1\leq l\leq N$ represents an elementary solution to (\ref{eq-2.1}) with q=q[0] at $\lambda=\lambda_l$. Then $N$-fold basic DT for Eq. (\ref{eq-1.2}) is thereby inferred that

\begin{equation}\label{eq-2.3}
\begin{split}
  \Psi[N] &= T[N]T[N-1]T[N-2]\cdots T[1]\Psi,~~T[l] = \lambda I-H[l]\Lambda_lH[l]^{-1},\\
  q[N] &= q[N-1]+2(\lambda_N^*-\lambda_N)\frac{\psi_N[N-1]^{*}\phi_N[N-1]}{(|\psi_N[N-1]|^2+|\phi_N[N-1]|^2)},
\end{split}
\end{equation}
where $\Psi_l[l-1]=(\psi_l[l-1],\phi_l[l-1])^{\dag}$, and
\begin{equation*}
\begin{split}
    \Psi_l[l-1] &= T_l[l-1]T_l[l-2]T_l[l-3]\cdots T_l[1]\Psi_l,\\
    T_l[k]& = T[k]|_{\lambda=\lambda_l}, 1\leq l\leq N,1\leq k\leq l-1,\\
  H[l-1] &=
  \begin{pmatrix}
   \psi_l[l-1] & -\phi_l[l-1]^{*}  \\
    \phi_l[l-1] & \psi_l[l-1]^{*}
    \end{pmatrix},~~
  \Lambda_l=
  \begin{pmatrix}
   \lambda_l & 0  \\
    0 & \lambda_l^{*}
    \end{pmatrix}.\\
\end{split}
\end{equation*}

Considering an elementary solution cannot be iterated many times by the above method, it is necessary to construct the generalized DT to overtake this difficulty. Therefore, suppose that $\Psi_1=\Psi_1(\lambda_1+\epsilon)$, which is a special solution of (\ref{eq-2.1}) with q[0] at $\lambda=\lambda_1+\epsilon$, applying the taylor expansion on $\Psi_1$ at $\epsilon=0$ yields
\begin{equation}\label{eq-2.4}
  \Psi_1=\Psi_1^{[0]}+\Psi_1^{[1]}\epsilon+\Psi_1^{[2]}\epsilon^2+\Psi_1^{[3]}\epsilon^3+\cdots+\Psi_1^{[N]}\epsilon^N+\cdots,
\end{equation}
with $\epsilon$ a small parameter and $\Psi_1^{[k]}=\frac{1}{k!}\frac{\partial^k}{\partial\lambda^k}\Psi_1(\lambda)|_{\lambda=\lambda_1}$. It is obvious that $\Psi_1^{[0]}$ is the solution of (\ref{eq-2.1}) with q=q[0] at $\lambda=\lambda_1$.

\subsection{The 1-fold generalized DT}

According to the basic DT (\ref{eq-2.2}), we can easily derive the 1-fold generalized DT formulas, that is

\begin{equation}\label{eq-2.5}
\begin{split}
  \Psi[1] &= T[1]\Psi,~~T[1]=\lambda I-H[0]\Lambda_1H[0]^{-1},\\
  q[1] &= q[0]+2(\lambda_1^{*}-\lambda_1)\frac{\psi_1[0]^{*}\phi_1[0]}{(|\psi_1[0]|^2+|\phi_1[0]|^2)},
\end{split}
\end{equation}
with $\phi_1[0]=\phi_1^{[0]}, \psi_1[0]=\psi_1^{[0]}$, and

\begin{equation*}
\begin{gathered}
  H[0]=
  \begin{pmatrix}
   \psi_1[0] & -\phi_1[0]^{*}  \\
    \phi_1[0] & \psi_1[0]^{*}
    \end{pmatrix},~~
  \Lambda_1=
  \begin{pmatrix}
   \lambda_1 & 0  \\
    0 & \lambda_1^{*}
    \end{pmatrix}.
\end{gathered}
\end{equation*}

\subsection{The 2-fold generalized DT}\label{sec-1.1}

Apparently, $T[1]\Psi_1$ is the solution of (\ref{eq-2.1}) with $q[1]$ at $\lambda=\lambda_1+\epsilon$ and $T_1[1]\Psi_1^{[0]}=0$. It is natural to draw the following result

\begin{equation*}
\begin{split}
  \lim_{\epsilon \to 0}\frac{T[1]|_{\lambda=\lambda_1+\epsilon}\Psi_1}{\epsilon}
  & = \lim_{\epsilon \to 0}\frac{(\epsilon+T_1[1])\Psi_1}{\epsilon}\\
  & = \Psi_1^{[0]}+T_1[1]\Psi_1^{[1]}\\
  & \equiv \Psi_1[1],\\
\end{split}
\end{equation*}
it gives a nonzero solution of the system (\ref{eq-2.1}) with q[1] at $\lambda=\lambda_1$. Hence, 2-fold generalized DT can be constructed, namely

\begin{equation}\label{eq-2.6}
  \begin{split}
    \Psi[2]& = T[2]T[1]\Psi,~~T[2] = \lambda I-H[1]\Lambda_2H[1]^{-1},\\
    q[2] &= q[1]+2(\lambda_1^{*}-\lambda_1)\frac{\psi_1[1]^{*}\phi_1[1]}{(|\psi_1[1]|^2+|\phi_1[1]|^2)},
  \end{split}
\end{equation}
with $\Psi_1[1]=(\psi_1[1],\phi_1[1])^{\dag}$, and
\begin{equation*}
\begin{gathered}
  H[1]=
  \begin{pmatrix}
   \psi_1[1] & -\phi_1[1]^{*}  \\
    \phi_1[1] & \psi_1[1]^{*}
    \end{pmatrix},~~
  \Lambda_2=
  \begin{pmatrix}
   \lambda_1 & 0  \\
    0 & \lambda_1^{*}
    \end{pmatrix}.
\end{gathered}
\end{equation*}

\subsection{The 3-fold generalized DT}

Continuing the similar process above, we give 3-fold generalized DT. Under the following conditions
\begin{equation*}
  T_1[1]\Psi_1^{[0]}=0,~~T_1[2](\Psi_1^{[0]}+T_1[1]\Psi_1^{[1]})=0,
\end{equation*}
and applying the limit method,
\begin{equation*}
\begin{split}
  \lim_{\epsilon \to 0}\frac{[T[2]T[1]]|_{\lambda=\lambda_1+\epsilon}\Psi_1}{\epsilon^2}
  & = \lim_{\epsilon \to 0}\frac{(T_1[2]+\epsilon)(T_1[1]+\epsilon)\Psi_1}{\epsilon^2}\\
  & = \Psi_1^{[0]}+(T_1[1]+T_1[2])\Psi_1^{[1]}+T_1[2]T_1[1]\Psi_1^{[2]}\\
  & \equiv \Psi_1[2],\\
  \end{split}
\end{equation*}
thus a nontrivial solution can be obtained for the Lax pair (\ref{eq-2.1}) with $q[2]$ at $\lambda=\lambda_1$. Then the 3-fold generalized DT is naturally deduced as follows

\begin{equation}\label{eq-2.7}
    \begin{split}
  \Psi[3] &= T[3]T[2]T[1]\Psi,~~T[3] = \lambda I-H[2]\Lambda_3H[2]^{-1},\\
  q[3] &= q[2]+2(\lambda_1^{*}-\lambda_1)\frac{\psi_1[2]^{*}\phi_1[2]}{(|\psi_1[2]|^2+|\phi_1[2]|^2)},
    \end{split}
\end{equation}
where $\Psi_1[2]=(\psi_1[2],\phi_1[2])^{\dag}$, and

\begin{equation*}
\begin{gathered}
  H[2]=
  \begin{pmatrix}
   \psi_1[2] & -\phi_1[2]^{*}  \\
    \phi_1[2] & \psi_1[2]^{*}
    \end{pmatrix},~~
  \Lambda_3=
  \begin{pmatrix}
   \lambda_1 & 0  \\
    0 & \lambda_1^{*}
    \end{pmatrix}.
\end{gathered}
\end{equation*}

\subsection{The N-fold generalized DT}

Repeating $N$ times of the above process, it naturally gives rise to the expression of $N$-fold generalized DT, which reads

\begin{equation}\label{eq-2.8}
\begin{split}
  \Psi_1[N-1] &= \Psi_1^{[0]}+\sum_{l=1}^{N-1}T_1[l]\Psi_1^{[1]}+\sum_{k=1}^{l-1}\sum_{l=1}^{N-1}T_1[l]T_1[k]\Psi_1^{[2]}
  +\cdots+T_1[N-1]T_1[N-2]\cdots T_1[1]\Psi_1^{[N-1]},\\
  \Psi[N] &= T[N]T[N-1]T[N-2]\cdots T[1]\Psi,~~T[N] = \lambda I-H[N-1]\Lambda_NH[N-1]^{-1},\\
  q[N] &= q[N-1]+2(\lambda_1^{*}-\lambda_1)\frac{\psi_1[N-1]^{*}\phi_1[N-1]}{(|\psi_1[N-1]|^2+|\phi_1[N-1]|^2)},
\end{split}
\end{equation}
where $\Psi_1[N-1]=(\psi_1[N-1],\phi_1[N-1])^{\dag}$, and

\begin{equation*}
\begin{gathered}
  H[l-1]=
  \begin{pmatrix}
   \psi_1[l-1] & -\phi_1[l-1]^{*}  \\
    \phi_1[l-1] & \psi_1[l-1]^{*}
    \end{pmatrix},~~
  \Lambda_l=
  \begin{pmatrix}
   \lambda_1 & 0  \\
    0 & \lambda_1^{*}
    \end{pmatrix},
    1\leq l \leq N.
\end{gathered}
\end{equation*}

Combining and simplifying the above formulas (\ref{eq-2.5}-\ref{eq-2.8}), it follows a compact formula for $N$-order rational solution of (\ref{eq-1.2}) that

\begin{equation}\label{eq-2.9}
  q[N]=q[0]+2(\lambda_1^{*}-\lambda_1)\sum_{j=0}^{N-1}\frac{\psi_1[j]^{*}\phi_1[j]}{(|\psi_1[j]|^2+|\phi_1[j]|^2)}.
\end{equation}

In the following section,  we can utilize the above formula to derive an arbitrary order rogue wave for (\ref{eq-1.2}). Then it follows their dynamic behavior illustrations.


\section{Rogue wave solutions}

Having established the result of generalized DT, attention is now given to constructing higher-order rogue waves for (\ref{eq-1.2}). For this purpose, assuming the seed solution
\begin{equation}\label{eq-3.1}
  q[0]=e^{i\theta},~~\theta=at+(-a^6\delta_6+30a^4\delta_6-90a^2\delta_6-\frac{1}{2}a^2+20\delta_6+1)z,~~a\in \mathbb{R}.
\end{equation}
The problem that a seed solution cannot be iterated by basic DT, can be solved by constructing its generalized DT.  And then substituting Eq. (\ref{eq-3.1}) into Eq. (\ref{eq-2.1}), the corresponding fundamental vector solution can be obtained, that is
\begin{equation}\label{eq-3.2}
  \Psi_1=
  \begin{pmatrix}
    i(C_1e^{M}-C_2e^{-M})e^{-\frac{1}{2}\theta} \\
    (C_2e^{M}-C_1e^{-M})e^{\frac{1}{2}\theta}
    \end{pmatrix},
\end{equation}
with
\begin{equation*}
\begin{split}
  C_1 =& \frac{(a+2\lambda+\sqrt{(a+2\lambda)^2+4})^{\frac{1}{2}}}{\sqrt{(a+2\lambda)^2+4}},~~
  C_2 =\frac{(a+2\lambda+\sqrt{(a+2\lambda)^2+4})^{\frac{1}{2}}}{\sqrt{(a+2\lambda)^2+4}},\\
  M =& \frac{1}{4}\sqrt{(a+2\lambda)^2+4} \Big\{[i(-2a^5+4a^4\lambda-8a^3\lambda^2+16a^2\lambda^3
    -32a\lambda^4+64\lambda^5+40a^3-48a^2\lambda\\
    & +48a\lambda^2-32\lambda^3-60a+24\lambda)\delta_6  +i(-a+2\lambda)]z+2it+\sum_{k=1}^Ns_k\xi^{2k}\Big\},~~s_k=m_k+in_k,(m_k,n_k\in\mathbb{R}),\\
\end{split}
\end{equation*}
where $\xi$ is a small real parameter. Setting $\lambda=-\frac{1}{2}a+i+\xi^2$ and expanding $\Psi_1$ at $\xi=0$, it arrives at
\begin{equation}\label{eq-3.3}
  \Psi_1(\xi)=\Psi_1^{[0]}+\Psi_1^{[1]}\xi^2+\Psi_1^{[2]}\xi^4+\cdots.
\end{equation}
Here, vector function $\Psi_1^{[0]}$  has the following explicit expression
\begin{equation}\label{eq-3.4}
  \psi_1^{[0]}=\frac{1+i}{2}\eta_1^{[0]}e^{-\frac{i}{2}\theta},~~\phi_1^{[0]}=-\frac{1+i}{2}\eta_2^{[0]}e^{\frac{i}{2}\theta},
\end{equation}
where
\begin{equation*}
  \begin{split}
    \eta_1^{[0]} &= 12(a^5-5ia^4-20a^3+30ia^2)\delta_6+2(180\delta_6+1)a-2i(60\delta_6+1),\\
    \eta_2^{[0]} &= 12(ia^5+5 a^4-20ia^3-30a^2)\delta_6+2i(180\delta_6+1)a+2(60\delta_6+1).
  \end{split}
\end{equation*}
Clearly, $\Psi_1^{[0]}=(\psi_1^{[0]},\phi_1^{[0]})^{\dag}$  satisfies the system (\ref{eq-2.1}) with spectral parameter $\lambda_1=-\frac{a}{2}+\frac{i}{2}$. Therefore, utilizing the formula (\ref{eq-2.9}) with $N=1$, it suffices to obtain first-order rogue wave of (\ref{eq-1.2}), that is

\begin{equation}\label{eq-3.5}
  q[1]=\Big{(}1+\frac{D_1+iE_1}{F_1}\Big{)}e^{i\theta},
\end{equation}
where
\begin{equation*}
\begin{split}
    F_1 = & 144(a^{10}-15a^8+160a^6-200a^4+300a^2+100)z^2\delta_6^2+48(a^6-15a^4+10)z^2\delta_6\\
    & -48(a^4-20a^2+30)azt\delta_6+4(a^2+1)z^2-8azt+4t^2+1,   \\
    D_1 = & -288(a^{10}-15a^8+160a^6-200a^4+300a^2+100)z^2\delta_6^2-96(a^6-15a^4+10)z^2\delta_6\\
    & +96(a^4-20a^2+30)azt\delta_6-8(a^2+1)z^2+16azt-8t^2+2, \\
    E_1 = & 240(a^4-6a^2+2)z\delta_6+8z.
\end{split}
\end{equation*}

\begin{figure*}[!htbp]
\centering
\subfigure[]{\includegraphics[height=1.5in,width=1.9in]{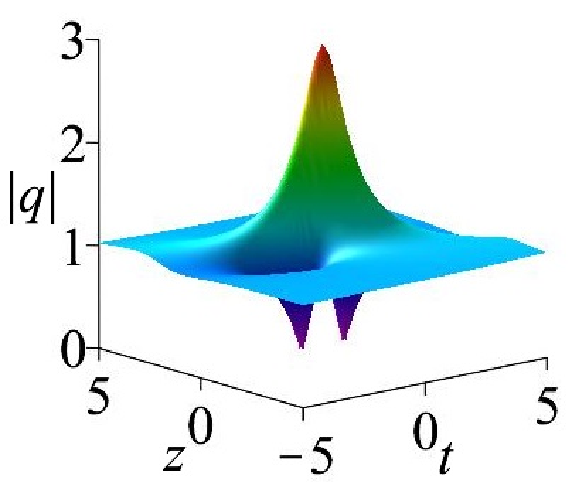}}
\subfigure[]{\includegraphics[height=1.5in,width=1.9in]{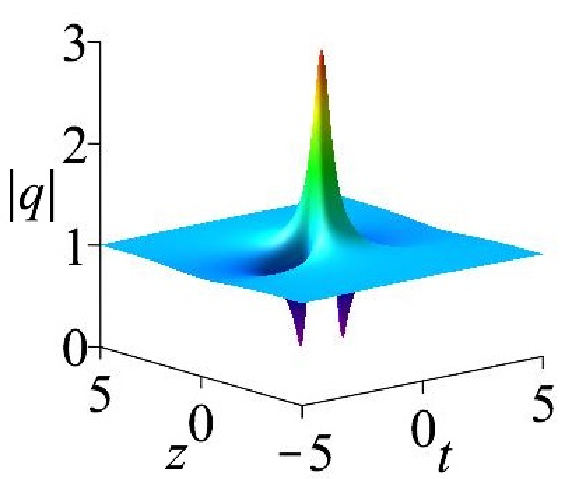}}
\subfigure[]{\includegraphics[height=1.5in,width=1.9in]{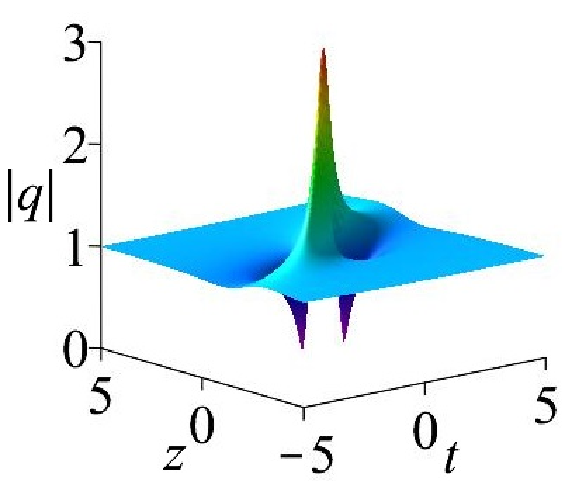}}
\caption{ Plots of first-order rogue wave with $a=-0.5, 0, 0.5$, from left to right, respectively and $\delta_6=0.01$. \label{Fig-3.1}}
\end{figure*}

\begin{figure*}[!htbp]
\centering
\subfigure[]{\includegraphics[height=1.3in,width=1.3in]{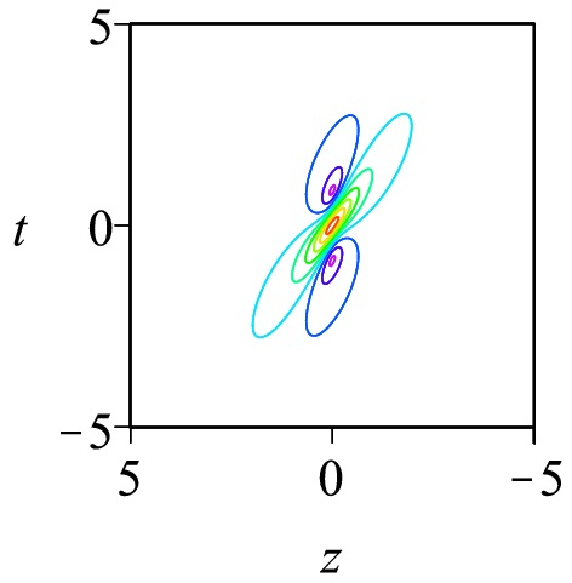}}
\subfigure[]{\includegraphics[height=1.3in,width=1.3in]{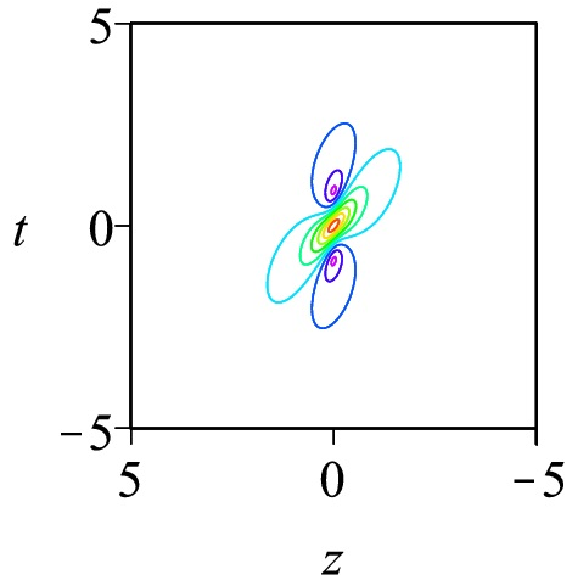}}
\subfigure[]{\includegraphics[height=1.3in,width=1.3in]{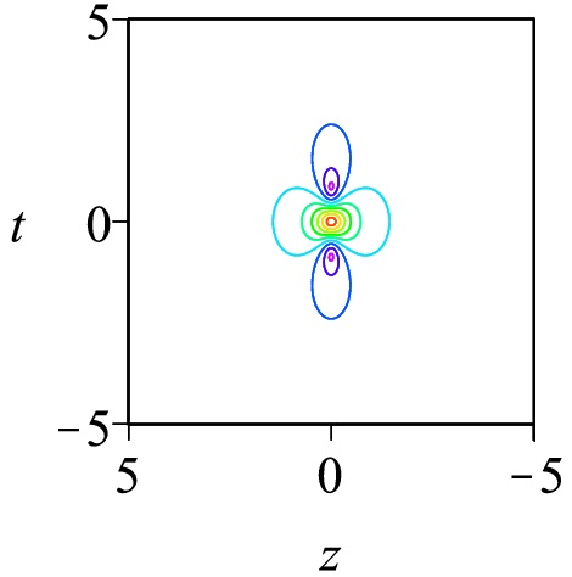}}
\subfigure[]{\includegraphics[height=1.3in,width=1.3in]{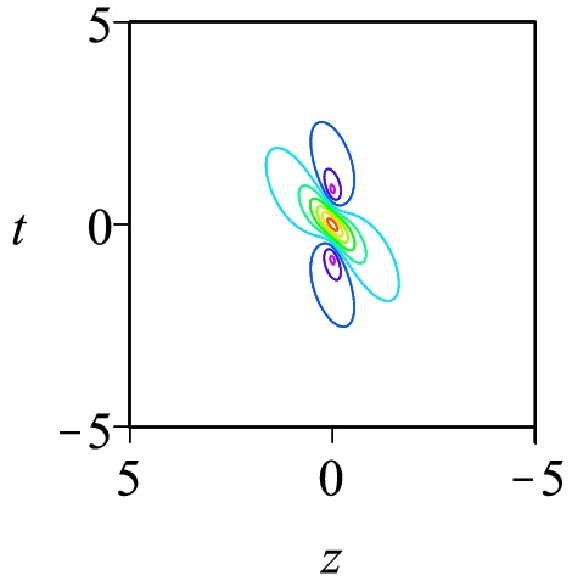}}
\subfigure[]{\includegraphics[height=1.3in,width=1.3in]{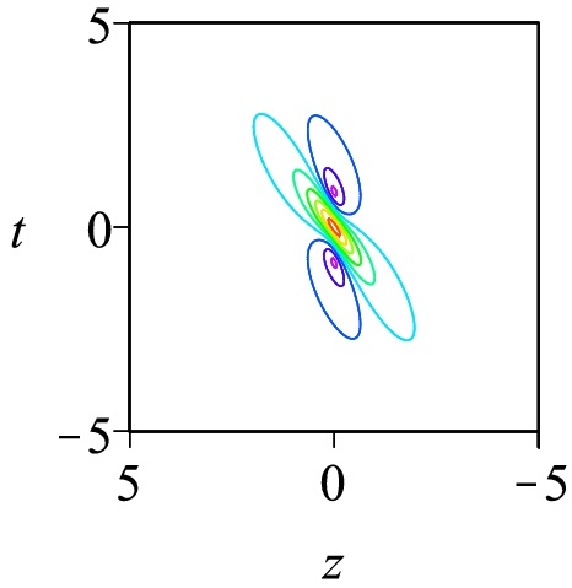}}
\caption{ Contour graphics for first-order rogue wave with $a=-0.5, -0.3, 0, 0.3, 0.5$, from left to right, respectively and $\delta_6=0.01$. \label{Fig-3.2}}
\end{figure*}

\begin{figure*}[!htbp]
\centering
\includegraphics[height=1.4in,width=6.4in]{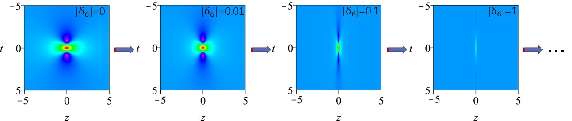}
\caption{ The evolution process of the rogue wave structure with the change of $\delta_6$ for $a=0$. \label{Fig-3.3}}
\end{figure*}

Obviously, there are two arbitrary parameters $a$ and $\delta_6$ in the expression for $q[1]$, the latter is the sixth-order dispersion coefficient. Next, we fix $\delta_6$ to analyze the dynamic of rogue wave solution with changing of frequency $a$.  Taking the case of $a=0$ as a criterion, when $a<0$, the crest of rogue wave occurs counterclockwise deflection; while $a>0$, the crest occurs clockwise deflection. In addition, the width of the crest changes. As $|a|$ increases, the deflection angle of crest of rogue wave increases, and so does its width. Figs. \ref{Fig-3.1} and \ref{Fig-3.2} illustrate the above dynamic characteristics. Now, we fix $a$ to be any particular constant and take limit on $q[1]$ at $\delta_6 \to \infty$, that is

\begin{equation}\label{eq-3.6}
  \lim_{|\delta_6| \to \infty}|q[1]|\equiv1.
\end{equation}
As the absolute value of $\delta_6$ increases, the modulus of $q[1]$ gradually reverts to a constant background plane; the rogue wave gradually disappeared and the energy gradually decreased. Without loss of generality, fix $a=0$, Fig. \ref{Fig-3.3} shows this evolution process in the first-order rogue wave structure by varing parameter $\delta_6$. For $\delta_6=0$, Eq.(\ref{eq-1.1}) degenerates into the standard NLS equation, and it follows that the amplitude of $|q[1]|$ is equal to 3.

\begin{figure*}[!htbp]
\centering
\subfigure[]{\includegraphics[height=1.5in,width=1.9in]{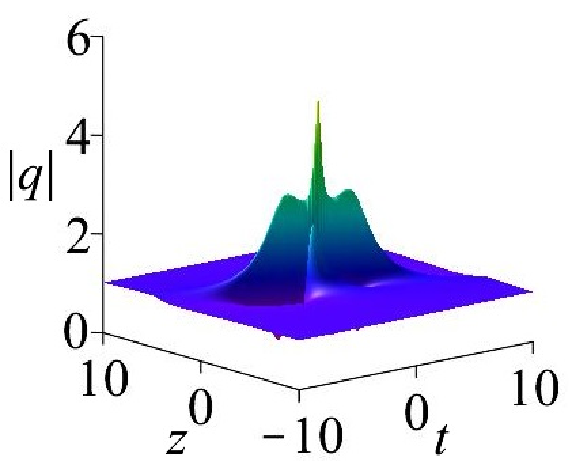}}\hspace{0.5cm}
\subfigure[]{\includegraphics[height=1.5in,width=1.9in]{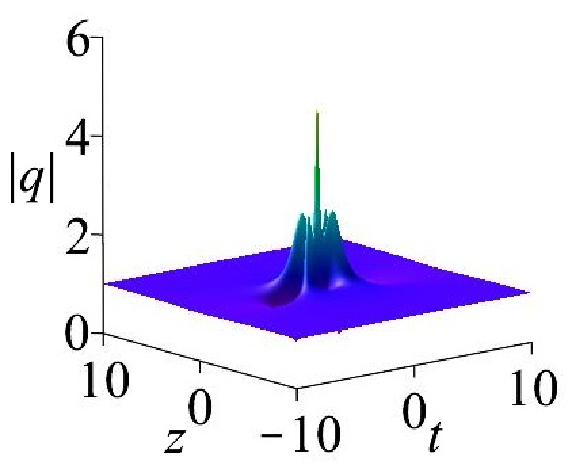}}\hspace{0.5cm}
\subfigure[]{\includegraphics[height=1.5in,width=1.9in]{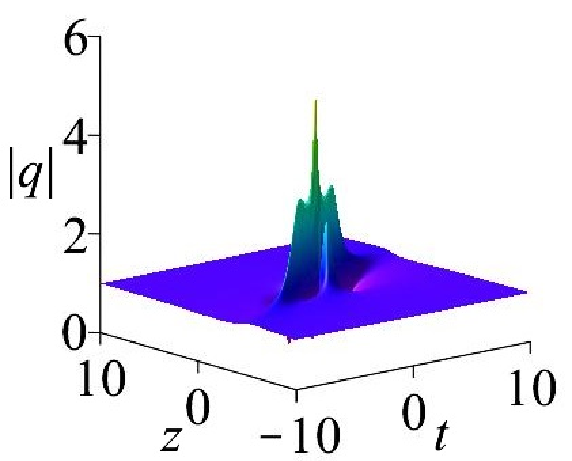}}
\caption{  Plots of second-order rogue wave by choosing $m_1=n_1=0$ and $a=-0.5, 0, 0.5$, from left to right, respectively. \label{Fig-3.4} }
\end{figure*}

\begin{figure*}[!htbp]
\centering
\subfigure[]{\includegraphics[height=1.5in,width=1.9in]{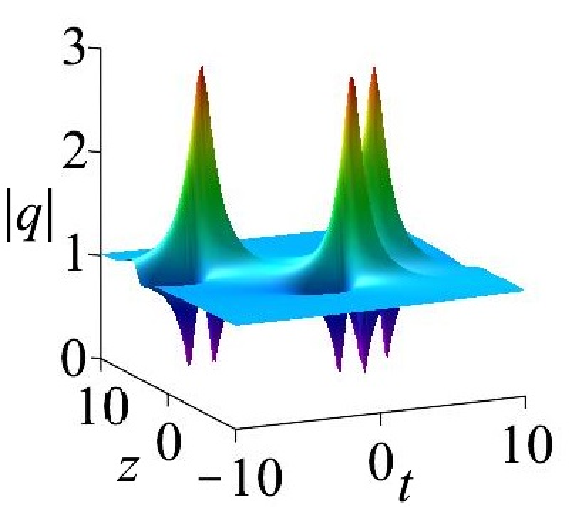}}
\subfigure[]{\includegraphics[height=1.5in,width=1.9in]{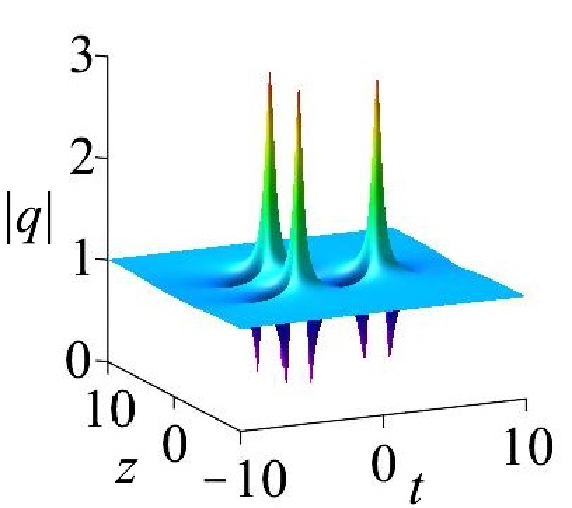}}
\subfigure[]{\includegraphics[height=1.5in,width=1.9in]{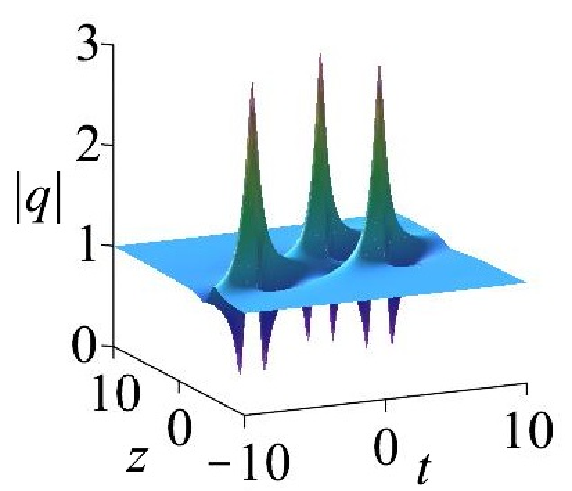}}
\caption{ 3D graphics for second-order rogue wave by choosing $\delta_6=0.01$, $m_1=100$, $n_1=0$ and $a=-0.5$, $0$, $0.5$, from left to right, respectively. \label{Fig-3.51}}
\end{figure*}

\begin{figure*}[!htbp]
\centering
\subfigure[]{\includegraphics[height=1.3in,width=1.3in]{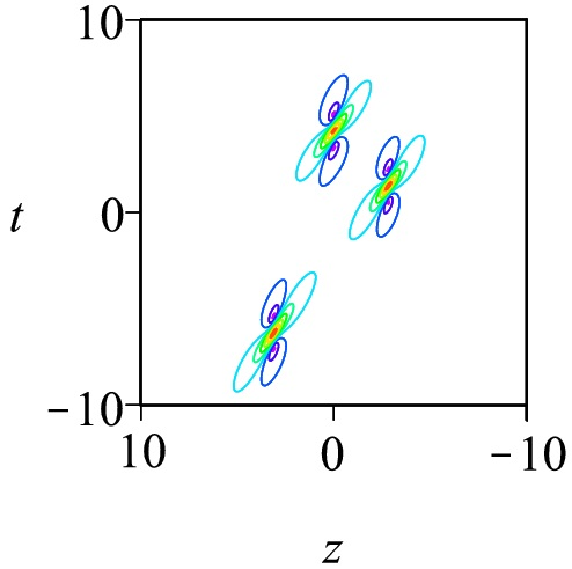}}
\subfigure[]{\includegraphics[height=1.3in,width=1.3in]{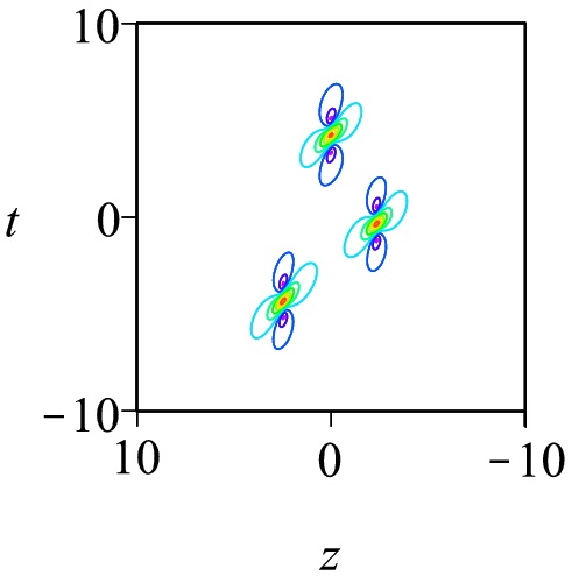}}
\subfigure[]{\includegraphics[height=1.3in,width=1.3in]{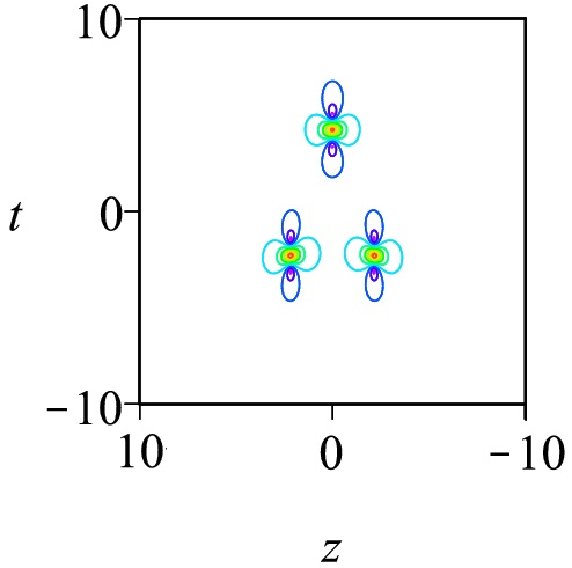}}
\subfigure[]{\includegraphics[height=1.3in,width=1.3in]{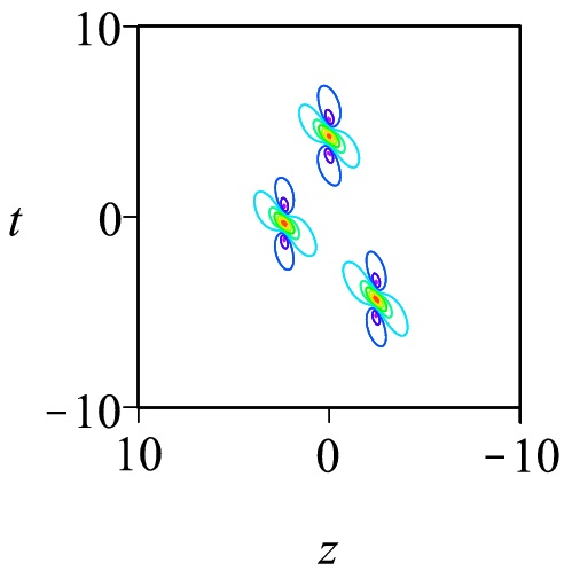}}
\subfigure[]{\includegraphics[height=1.3in,width=1.3in]{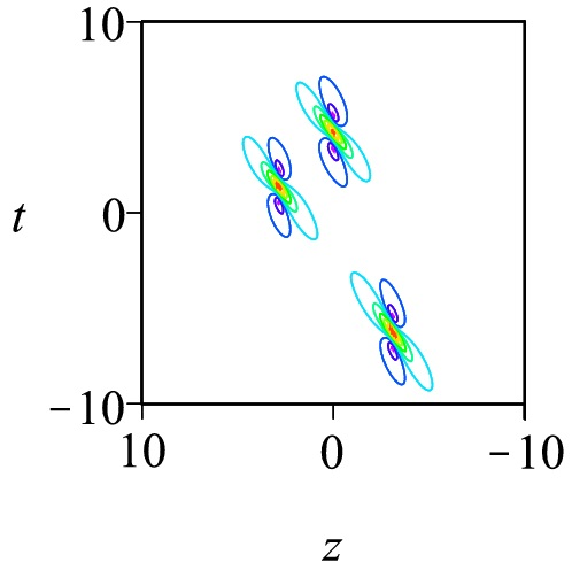}}
\caption{ The corresponding contour graphics for the second-order rogue wave obtained in Fig. \ref{Fig-3.51} with parameters: $\delta_6=0.01$, $m_1=100$, $n_1=0$ and $a=-0.5$, $-0.3$, $0$,$0.3$, $0.5$, from left to right, respectively. \label{Fig-3.5}}
\end{figure*}

Similar to the computational process of Section \ref{sec-1.1}, taking limit

\begin{equation}\label{eq-3.7}
\begin{split}
  \lim_{\xi \to 0}\frac{T[1]|_{\xi=-\frac{1}{2}a+i+\xi^2}\Psi_1}{\xi^2}
  &= \lim_{\xi \to 0}\frac{(\xi^2+T_1[1])\Psi_1}{\xi^2}\\
  &= \Psi_1^{[0]}+T_1[1]\Psi_1^{[1]}\\
  & \equiv \Psi_1[1],
  \end{split}
\end{equation}
and using the obtained formula (\ref{eq-2.9}) with $N=2$, it is not hard to adduce the second-order rogue wave. Since the expression of this solution is too cumbersome, we only show its dynamic behavior, which are illustrated by Figs. \ref{Fig-3.4} and \ref{Fig-3.51}.  The corresponding contour map of Fig. \ref{Fig-3.51} is demonstrated in Fig. \ref{Fig-3.5}. Substituting $m_1=0, n_1=0$ into Eq. (\ref{eq-2.9}), the second-order fundamental rogue wave solution can be derived, and there exists a maximum value 5 at point $(0,0)$ in the $(t,z)$ plane, see Fig. \ref{Fig-3.4}. However, when only changing a parameter $m_1=100$, the fundamental structure disappears, there appears a triplet structure containing three first-order rogue waves. Similarly, the deflation properties in first-order rogue wave also exist in the above two kinds of second-order rogue wave structures . To see this, the evolution process of this corresponding rogue wave structure is demonstrated in Fig. \ref{Fig-3.5} with varying $a$.

\begin{figure*}[!htbp]
\centering
\subfigure[]{\includegraphics[height=1.5in,width=1.9in]{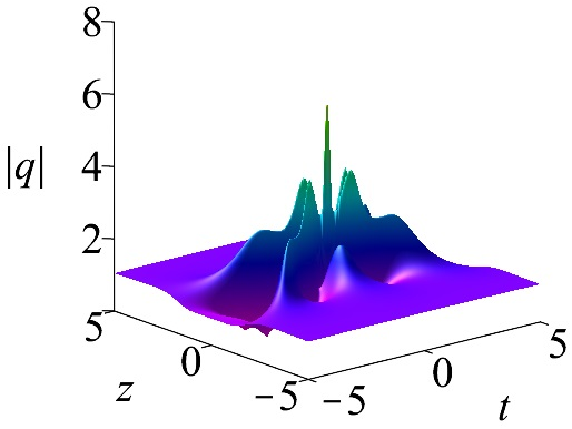}}\hspace{0.5cm}
\subfigure[]{\includegraphics[height=1.5in,width=1.9in]{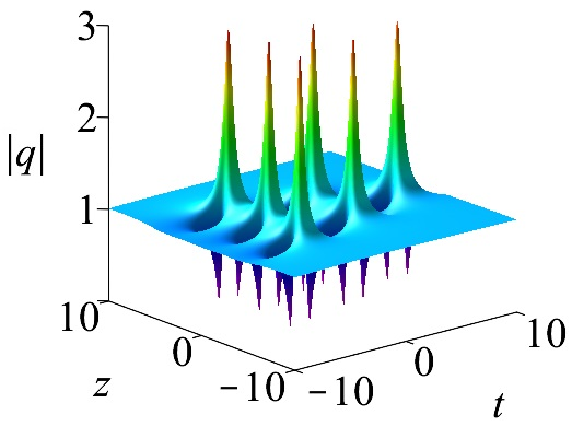}}\hspace{0.5cm}
\subfigure[]{\includegraphics[height=1.5in,width=1.9in]{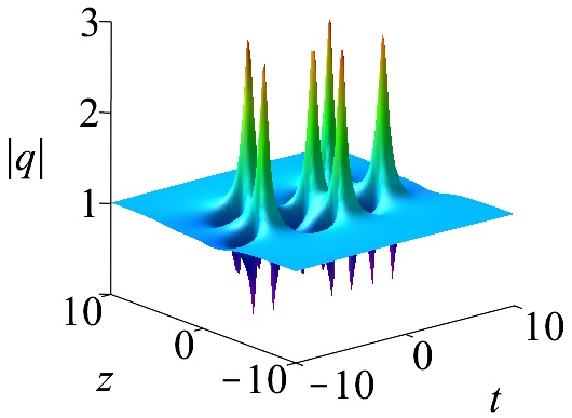}}\hspace{0.5cm}\\
\subfigure[]{\includegraphics[height=1.9in,width=1.9in]{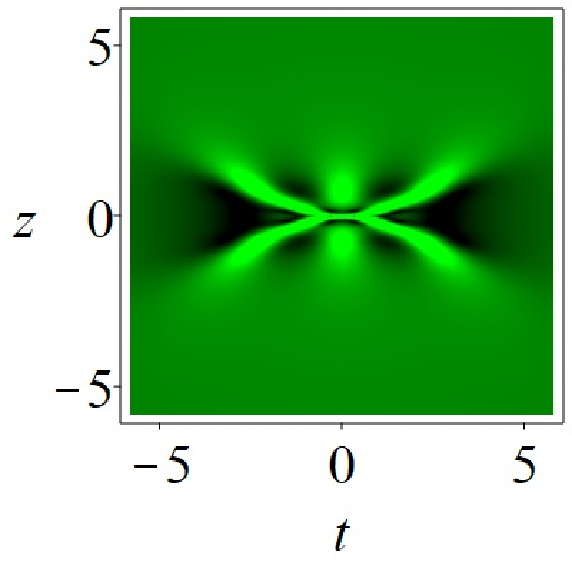}}\hspace{0.5cm}
\subfigure[]{\includegraphics[height=1.9in,width=1.9in]{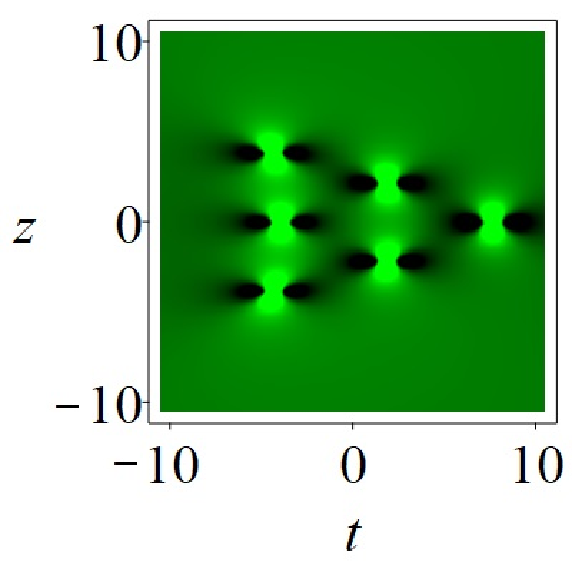}}\hspace{0.5cm}
\subfigure[]{\includegraphics[height=1.9in,width=1.9in]{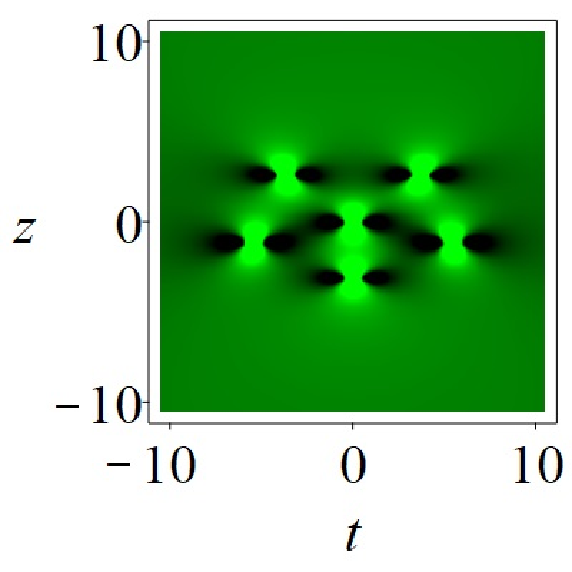}}
\caption{  Three kinds of third-order rogue wave structures for Eq. (\ref{eq-1.2}). Left columns: fundamental type structure at $a=0, \delta_6=0.01, m_i=n_i=0 (i=1, 2)$; middle columns: triangular structure at $m1=100$, the rest of the parameters are same to left columns; right columns: circular structure at $m_2=1000$, the rest of the parameters are same to left columns. \label{Fig-3.6} }
\end{figure*}

Applying formula (\ref{eq-2.9}) with $N=3$, it then follows the third-order rogue wave solution. Here, we just show three types of third-order rogue wave solutions, fundamental pattern, triangular pattern and circular pattern rogue wave, respectively, see Fig. \ref{Fig-3.6}. The first row are the three-dimensional graphs, and the second row are the corresponding density maps. The amplitude of third-order fundamental rogue wave reaches maximum value 7 at point $(0,0)$ in the $(t,z)$ plane. Obviously, these rogue waves are symmetrical, which can be seen from Figs. \ref{Fig-3.6}(d-f). They also possess the above deflection characteristics.

\section{Spectral analysis of rogue waves}

Our attention is now turned to spectral analysis on rogue wave solution for Eq. \eqref{eq-3.5} in this section. In \cite{akhmediev-pla-2011}, it appears that the specific triangular spectrum for a Peregrine rogue wave could be applied to early warning of rogue waves by spectral measurements. The spectrum analysis is referred to as a useful method in predicting and exciting of rogue wave solutions in the nonlinear fiber \cite{akhmediev-pd-2015,bayindir-pre-2016}. For more conveniently to calculate the spectral of first-order rogue wave solution, we take $\delta_6=\frac{1}{12}$ in Eq. \eqref{eq-3.5}. It then follows that
\begin{equation}\label{eq-5.1}
  q[1]=\Big{(}\frac{4(1+iK_1)}{K_2}-1\Big{)}
  \exp (i\theta_0),
\end{equation}
where
\begin{equation*}
\begin{split}
    K_1 = & (5a^4-30a^2+12)z, ~~~~\theta_0 =  at+\Big{(}\frac{5}{2}a^4-\frac{97}{12}a^2+\frac{8}{3}\Big{)}z,\\
    K_2 = & (a^{10}-15a^8+160a^6-260a^4+304a^2+144)z^2-4(a^4-20a^2+32)azt+4t^2+1.
\end{split}
\end{equation*}
Now we perform spectrum analysis approach on the above derived first-order rogue wave solution by the Fourier transformation as follows
\begin{equation}\label{eq-5.2}
  F(\beta,z)=\frac{1}{\sqrt{2\pi}}\int_{-\infty}^{+\infty}q[1](z,t)\exp(i\beta t)dt.
\end{equation}
From the solution \eqref{eq-5.1}, it is inferred that the rogue wave solution contains two parts, a plane  wave and a variable signal part. It is clear that the plane wave background becomes infinity and the integral is a $\delta$ function, so we omit the spectrum of plane wave background. The corresponding modulus of the rogue wave signal is given by
\begin{equation}\label{eq-5.3}
  |F(\beta,z)|=\sqrt{2\pi}\exp\Big{(}-\frac{|\beta'|}{2}\sqrt{1+(5a^4-30a^2+12)^2z^2}\Big{)},
\end{equation}
where $\beta'=\beta+a$.

\begin{figure*}[!htbp]
\centering
\subfigure[]{\includegraphics[height=1.9in,width=1.9in]{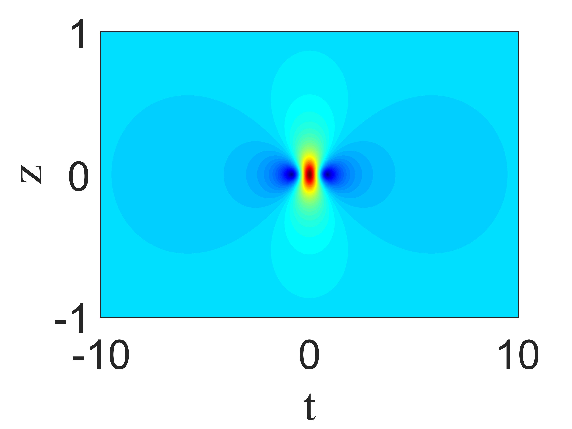}}\hspace{0.5cm}
\subfigure[]{\includegraphics[height=1.9in,width=1.9in]{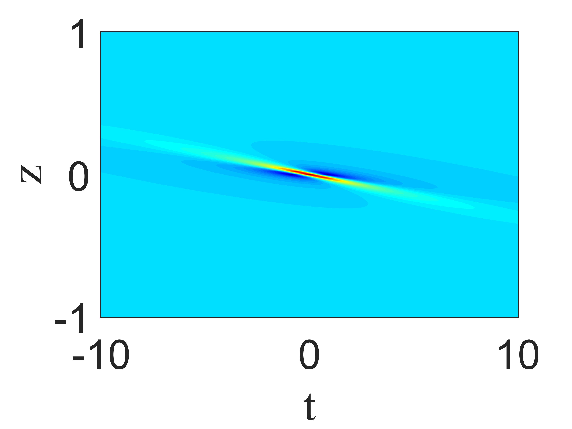}}\hspace{0.5cm}
\subfigure[]{\includegraphics[height=1.9in,width=1.9in]{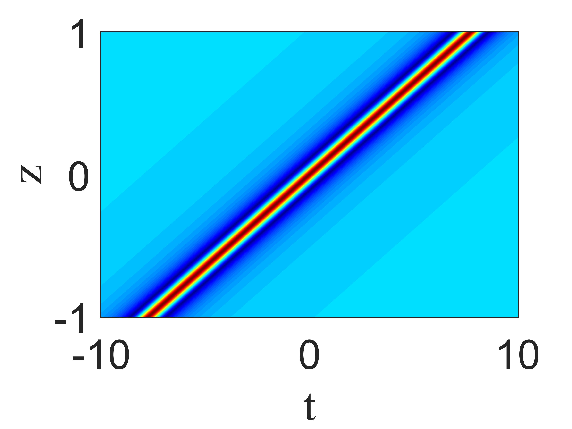}}\hspace{0.5cm}\\
\subfigure[]{\includegraphics[height=1.9in,width=1.9in]{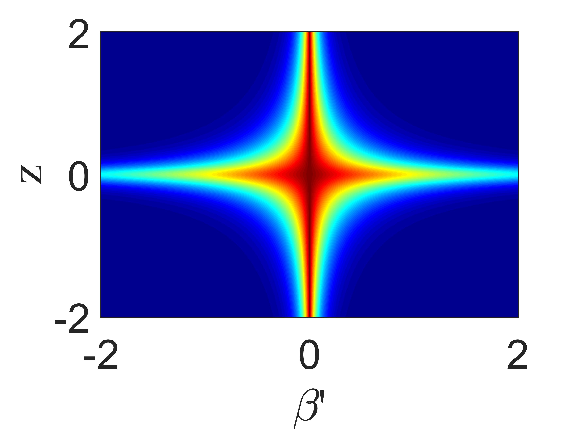}}\hspace{0.5cm}
\subfigure[]{\includegraphics[height=1.9in,width=1.9in]{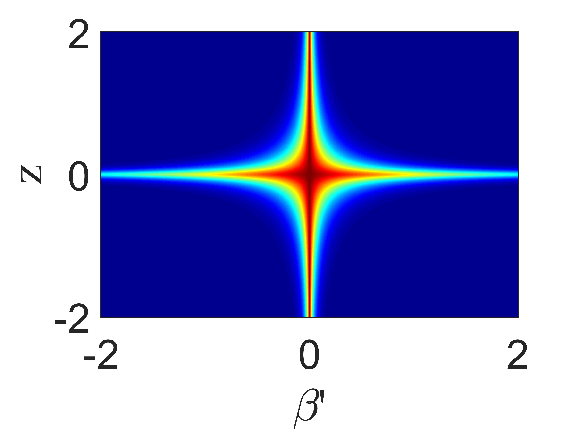}}\hspace{0.5cm}
\subfigure[]{\includegraphics[height=1.9in,width=1.9in]{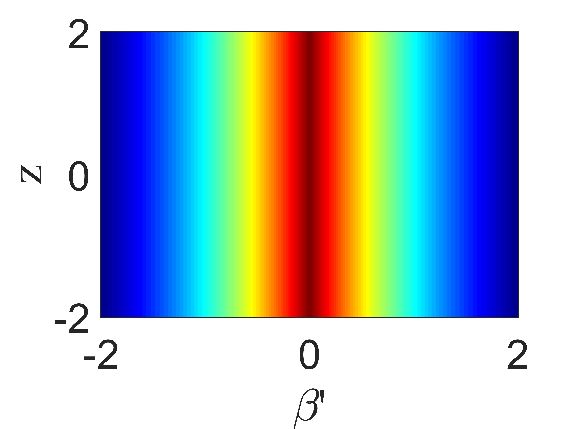}}
\caption{ The first row displays density figures of two first-order rogue waves and a W-shaped soliton solutions in Eq. \eqref{eq-5.1} with $a=0$, $2$, and $\frac{1}{5}\sqrt {75-\sqrt {165}}$, from left to right, respectively. The bottom row displays the spectrum dynamics of $|F(\beta,z)|$ in Eq.(\ref{eq-5.3}).  \label{Fig-5.1} }
\end{figure*}

Firstly, from the perspective of the bottom row of Fig. \ref{Fig-5.1} to analyze, it is clear that the spectrum of the solution \eqref{eq-5.1} with different $a$ has strong symmetry properties. And then combined with the expression \eqref{eq-5.3}, when $a\neq \pm \frac{1}{5} \sqrt{75\pm5\sqrt {165}}$, Figs. \ref{Fig-5.1}(d) and \ref{Fig-5.1}(e) have specific triangular spectrum of a Peregrine rogue wave. Furthermore, one can easily find that the triangular widening appears at $a=0$, when compared to the case at $a=2$. The corresponding density diagrams are displayed in Figs. \ref{Fig-5.1}(a) and \ref{Fig-5.1}(b). However, when $a=\pm\frac{1}{5}\sqrt {75\pm5\sqrt {165}}$, the spectrum of the solution \eqref{eq-5.1} is a band shape, see Fig. \ref{Fig-5.1}(f), and the rogue wave solution is now reduced to a stable W-shaped soliton solution in Fig. \ref{Fig-5.1}(c). Similarly, $N$th-order rogue waves can be reduced to $N$th-order W-shaped solitons. This result in turn implies that MI analysis is consistent with spectral analysis for the sixth-order Eq. (\ref{eq-1.2}) with $\delta_6=\frac{1}{12}$. From the perspective of MI gain function \eqref{eq-4.3}, it can be adduced that
\begin{equation}\label{eq-5.4g}
  g_6=\frac{1}{6}(30A^4-10\Omega^2A^2-90a^2A^2+\Omega^4+15a^2\Omega^2+15a^4+6),
\end{equation}
where $A$ is the amplitude, $\Omega$ is the perturbed frequency and $a$ is the frequency of background.
By setting $A=1, \Omega=0$ and $g_6=0$, it follows
\begin{equation}\label{eq-5.4g0}
  g_6=\frac{1}{2}(5a^4-30a^2+12)=0,
\end{equation}
then there is a transformation of two sates here, which happens between the rogue wave and the W-shaped soliton in the region of zero-frequency MS.

In order to demonstrate the impact of the parameter $\delta_6$, we will give the spectrum analysis of the rogue wave solution by selecting $a=0$ in Eq. \eqref{eq-3.5} for convenience. It thus transpires that
\begin{equation}\label{eq-5.4}
  q[1]=\Big{(}\frac{4(1+2i(1+60\delta_6)z)}{4t^2+4(1+60\delta_6)^2z^2+1}-1\Big{)}\exp (i(1+20\delta_6)z),
\end{equation}
and
\begin{equation}\label{eq-5.5}
  |F(\beta,z)|=\sqrt{2\pi}\exp\Big{(}-\frac{|\beta|}{2}\sqrt{1+4(1+60\delta_6)^2z^2}\Big{)}.
\end{equation}

\begin{figure*}[!htbp]
\centering
\subfigure[]{\includegraphics[height=1.9in,width=1.9in]{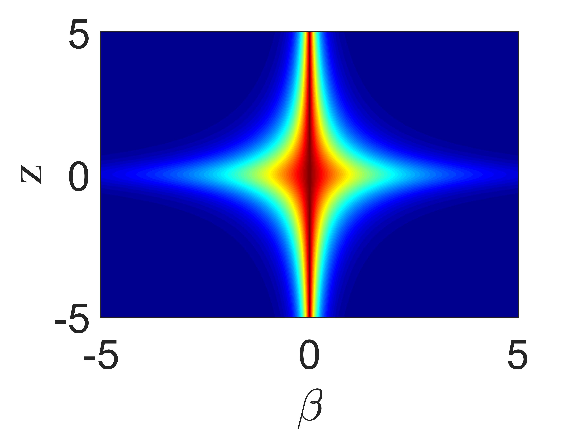}}\hspace{0.5cm}
\subfigure[]{\includegraphics[height=1.9in,width=1.9in]{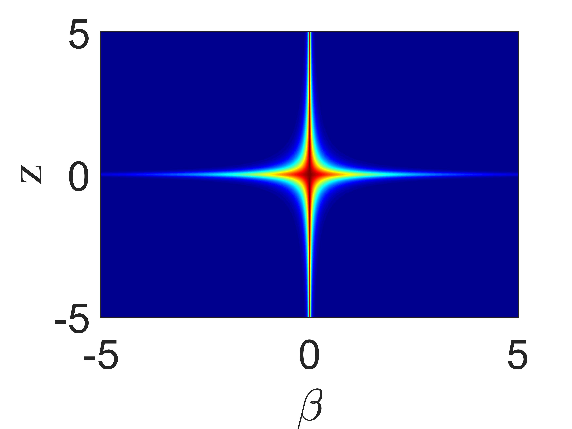}}\hspace{0.5cm}
\subfigure[]{\includegraphics[height=1.9in,width=1.9in]{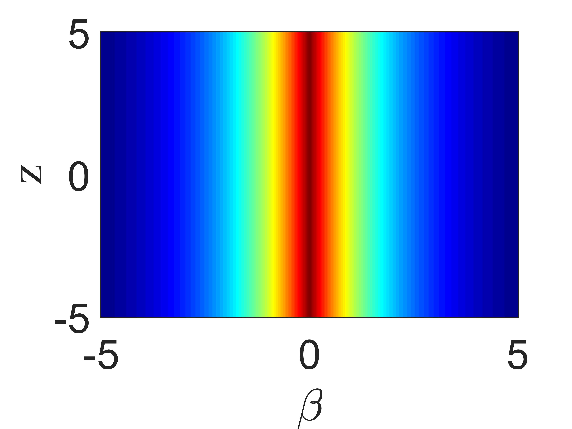}}
\caption{  Spectral dynamics of $|F(\beta,z)|$ in Eq. (\ref{eq-5.5}) with $\delta_6=0$, $0.1$ and $-\frac{1}{60}$, from left to right, respectively.  \label{Fig-5.2} }
\end{figure*}

Similarly, when $\delta_6\neq -\frac{1}{60}$, the spectrum of the solution \eqref{eq-5.4} also possesses specific triangular spectrum of a Peregrine rogue wave. In addition, their spectrum share the same features with different parameters as above Eq. \eqref{eq-5.1}. Obviously, a little change appears in $\delta_6$, a big change presents in the corresponding triangular widening spectrum in contrast with above condition. Setting $\delta_6=-\frac{1}{60}$, the solution \eqref{eq-5.4} is a W-shaped soliton, presented in Fig. \ref{Fig-5.2}(c). Its corresponding spectrum appears in banded form, see Fig. \ref{Fig-5.2}.

\section{Summary and discussions}

In conclusion, MI of the continuous wave background has been investigated for the NLS equation with different higher-order dispersion term. The MI distribution characteristics from the sixth-order to the eighth-order NLS equations are studied in detail. There are two arbitrary parameters, namely, higher-order dispersion term $\delta_i, i=3,4,\ldots$ and amplitude $A$. These parameters control the MS distribution of the NLS with different higher-order dispersion terms in the MI band. By adjusting the parameters, the MS quasi-elliptic and MS elliptic ring can be completely contained within the MI band or intersected at the MI boundary, the latter case yields two curves in the MI band. $g_i$ is a polynomial with $\omega$, and its highest power of $\omega$ is closely related to how many MS curves can exist in the MI band. It is adduced that the high-order dispersion terms indeed affect the distribution of the MS regime, $n$-order dispersion term corresponds to $n-2$ modulation stability curves in the MI band. Here, we do not consider the case of NLS equation with multiple dispersion terms, only considering the case with a higher-order dispersion term, it is inferred that the distribution of MS curve in the MI band is not affected. Therefore, when multiple different order dispersion terms exist simultaneously, the higher-order dispersion term plays the main role in the distribution of MS curve in the MI band.

In addition, we have proposed rogue waves of the NLS with sixth-order dispersion term. By constructing the generalized DT, we give compact algebraic expression of $N$th-order rogue waves (\ref{eq-2.9}). Then the exact expression of first-order rogue wave is demonstrated. Since expressions of high-order rogue waves are too cumbersome, we demonstrate its dynamic behavior through pictures. There are two arbitrary parameters $a$ and $\delta_6$, the sign of the former determines the direction of deflection and the magnitude of the absolute value affects the angle of deflection and the width of rogue wave solution. While the latter can cause the change of the width and amplitude of rogue wave. For the first- to third-order rogue waves, they all own the deflection properties mentioned above. For the third-order rogue wave solution, three kinds of structures, that is fundamental, triangular, and circular, are illustrated in Fig. \ref{Fig-3.6}, and their dynamic behavior features are discussed in detail.

Via the spectrum analysis approach on first-order rogue wave, It has been found that arbitrary parameters $a$ and $\delta_6$ have effects on the spectrum of the solution \eqref{eq-5.1}. Fixing $\delta_6=\frac{1}{12}$, when $a\neq\pm\frac{1}{5}\sqrt {75\pm5\sqrt {165}}$, the solution has specific triangular spectrum for a Peregrine rogue wave and the value of $a$ is related to the size of the triangular spectrum; when $a=\pm\frac{1}{5}\sqrt {75\pm5\sqrt {165}}$, the solution is reduced to a W-shaped soliton, which is not localized in  temporal and spatial context and the spectrum is banded. Similarly, fixing $a$, when $\delta_6$ satisfies certain constraint, the spectrum of the solution \eqref{eq-5.4} also presents the specific triangular spectrum or banded spectrum, which correspond to the rogue wave solution or W-shaped soliton solution, respectively.

Finally, it is worthy to mention that we will further study the excitation conditions and numerical analysis of various nonlinear waves and their corresponding positions in the MI gain plane in the future.

\section*{Acknowledgment}
The project is supported by the National Natural Science Foundation of China (Nos. 11675054 and 11435005), Global Change Research Program of China (No. 2015CB953904), and Shanghai Collaborative Innovation Center of
Trustworthy Software for Internet of Things (No. ZF1213).


\begin{thebibliography}{10}
%
\bibitem{draperl-mo-1965} Draper L, Freak wave. Marine Observer 35 (1965) 193-195.




\bibitem{bwj-2006-kc-2009} Kharif C, Pelinovsky E, Slyunyaev A, Quasi-linear wave focusing. Rogue waves in the ocean. Berlin: Springer (2009) 63-89.

\bibitem{akmhmedievn-pla-2009} Akhmediev N, Ankiewicz A, Taki M, Waves that appear from nowhere and disappear without a trace. Phys. Lett. A 373 (2009) 675-678.

\bibitem{zve-pd-2009} Zakharov V E, Ostrovsky L A, Modulation instability: the beginning. Physica D 238 (2009) 540-548.


\bibitem{raptiz-pre-2004} Rapti Z, Kevrekidis P G, Smerzi A, Bishop A R, Variational approach to the modulational instability. Phys. Rev. E 69 (2004) 017601.



\bibitem{bjf-1967} Benjamin T B, Feir J E, The disintegration of wave trains on deep water Part 1. Theory. J. Fluid Mech. 27 (1967) 417-430 .


\bibitem{taniutit-prl-1968} Taniuti T, Washimi H, Self-trapping and instability of hydromagnetic waves along the magnetic field in a cold plasma. Phys. Rev. Lett. 21 (1968) 209.



\bibitem{bespalovvi-1966} Bespalov V I, Talanov V I, Filamentary structure of light beams in nonlinear liquids. Zh. Eksp. Teor. Fiz. Pis'ma Red., 3 (1966) 471, English translation: JETP Lett. 3 (1966) 307.

\bibitem{withamgb-1965} Witham G B, Non-linear dispersive waves. Proc. R. Soc. Lond. A 283 (1965) 238-261.

\bibitem{zhaolc-pre-2014} Zhao L C, Xin G G, Yang Z Y, Rogue-wave pattern transition induced by relative frequency. Phys. Rev. E 90 (2014) 022918.

\bibitem{jhzhang-prsa-2017} Zhang J H, Wang L, Liu C, Superregular breathers, characteristics of nonlinear stage of modulation instability induced by higher-order effects. Proc. R. Soc. A 473 (2017) 20160681.

\bibitem{wangx-jmaa-2017} Wang X, Liu C, Wang L, Darboux transformation and rogue wave solutions for the variable-coefficients coupled Hirota equations, J. Math. Anal. Appl. 449 (2017) 1534-1552.

\bibitem{wangl-pre-2016} Wang L, Zhang J H, Wang Z Q, Liu C, Li M, Qi F H, Guo R, Breather-to-soliton transitions, nonlinear wave interactions, and modulational instability in a higher-order generalized nonlinear Schr\"{o}dinger equation. Phys. Rev. E 93 (2016) 012214.



\bibitem{gpagrawal-nfo-2006} Agrawal G P, Nonlinear Fiber Optics, Academic, New York, 2006.


\bibitem{stenflol-jpp-2010} Stenflo L, Marklund M, Rogue waves in the atmosphere. J. Plasma Phys. 76 (2010) 293-295.


\bibitem{aosborne-2010} Geist E L, Book review: Nonlinear ocean waves and the inverse scattering transform. 2011.



\bibitem{peregrinedh-jams-1983} Peregrine D H, Water waves, nonlinear Schr\"{o}dinger equations and their solutions. J. Aust. Math. Soc. 25 (1983) 16-43.

\bibitem{akhmediewn-pre-2009} Akhmediev N, Ankiewicz A, Soto-Crespo J M, Rogue waves and rational solutions of the nonlinear Schr\"{o}dinger equation. Phys. Rev. E 80 (2009) 026601.

\bibitem{Ankiewicza-jpagp-2010} Ankiewicz A, Clarkson P A, Akhmediev N, Rogue waves, rational solutions, the patterns of their zeros and integral relations. J. Phys. A: Math. Theor. 43 (2010) 122002.






\bibitem{ohtay-prsampes-2012} Ohta Y, Yang J K, General high-order rogue waves and their dynamics in the nonlinear Schr\"{o}dinger equation. Proc. R. Soc. A 468 (2012) 1716-1740.

\bibitem{zhanggq-prsa-2017} Zhang G Q, Yan Z Y, Wen X Y. Modulational instability, beak-shaped rogue waves, multi-dark-dark solitons and dynamics in pair-transition-coupled nonlinear Schr\"{o}dinger equations. Proc. R. Soc. A 473 (2017) 20170243.

\bibitem{gagnonl-jpa-1993} Gagnon L, Winternitz P, Symmetry classes of variable coefficient nonlinear Schr\"{o}dinger equations. J. Phys. A: Math. Gen. 26 (1993) 7061.


\bibitem{wangl-pre-2016-1} Wang L, Zhang J H, Liu C, Li M, Qi F H, Breather transition dynamics, Peregrine combs and walls, and modulation instability in a variable-coefficient nonlinear Schr\"{o}dinger equation with higher-order effects. Phys. Rev. E 93 (2016) 062217.



\bibitem{yangyq-nd-2015} Yang Y Q, Wang X, Yan Z Y, Optical temporal rogue waves in the generalized inhomogeneous nonlinear Schr\"{o}dinger equation with varying higher-order even and odd terms. Nonlinear Dyn. 81 (2015) 833-842.


\bibitem{zhanggq-cnsns-2018} Zhang G Q, Yan Z Y, Three-component nonlinear Schr\"{o}dinger equations: Modulational instability, Nth-order vector rational and semi-rational rogue waves, and dynamics. Commun. Nonlinear Sci. Numer. Simulat. 62 (2018) 117-133.

\bibitem{xut-nd-2017} Xu T, Chen Y, Mixed interactions of localized waves in the three-component coupled derivative nonlinear Schr\"{o}dinger equations. Nonlinear Dyn. 92 (2018) 2133-2142.


\bibitem{zhanggq-prsa-2018} Zhang G Q, Yan Z Y, The n-component nonlinear Schr\"{o}dinger equations: dark-bright mixed N-and high-order solitons and breathers, and dynamics. Proc. R. Soc. A 474 (2018) 20170688.


\bibitem{yangb-nd-2018} Yang B, Chen Y, Dynamics of high-order solitons in the nonlocal nonlinear Schr\"{o}dinger equations. Nonlinear Dyn. 94 (2018) 489-502.




\bibitem{ankiewicza-pre-2014} Ankiewicz A, Wang Y, Wabnitz S, Akhmediev N, Extended nonlinear Schr\"{o}dinger equation with higher-order odd and even terms and its rogue wave solutions. Phys. Rev. E 89 (2014) 012907.

\bibitem{caily-nd-2017} Cai L Y, Wang X, Wang L, Li M, Liu Y, Shi Y Y, Nonautonomous multi-peak solitons and modulation instability for a variable-coefficient nonlinear Schr\"{o}dinger equation with higher-order effects. Nonlinear Dyn. 90 (2017) 2221-2230.

\bibitem{berge-pr-1998} Berg\'{e} L, Wave collapse in physics: principles and applications to light and plasma waves. Physics reports 303 (1998) 259-370.

\bibitem{ankiewicza-pre-2016} Ankiewicz A, Kedziora D J, Chowdury A, Bandelow U, Akhmediev N, Infinite hierarchy of nonlinear Schr\"{o}dinger equations and their solutions. Phys. Rev. E 93 (2016) 012206.


\bibitem{hirotar-jmp-1973} Hirota R, Exact envelope-soliton solutions of a nonlinear wave equation. J. Math. Phys. 14 (1973) 805-809.


\bibitem{lakshmananm-pla-1988} Lakshmanan M, Porsezian K, Daniel M, Effect of discreteness on the continuum limit of the Heisenberg spin chain. Phys. Lett. A 133 (1988) 483-488.

\bibitem{chowdurya-pre-2014} Chowdury A, Kedziora D J, Ankiewicz A, Akhmediev N, Soliton solutions of an integrable nonlinear Schr\"{o}dinger equation with quintic terms. Phys. Rev. E 90 (2014) 032922.

\bibitem{sunjj-epjp-2017} Su J J, Gao Y T. Bilinear forms and solitons for a generalized sixth-order nonlinear Schr\"{o}dinger equation in an optical fiber. Eur. Phys. J. Plus 132 (2017) 53.

\bibitem{sunwr-adp-2017} Sun W R. Breather-to-soliton transitions and nonlinear wave interactions for the nonlinear Schr\"{o}dinger equation with the sextic operators in optical fibers. Ann. Phys. 529 (2017) 1600227.

\bibitem{lanzz-oqe-2018} Lan Z Z, Guo B L, Conservation laws, modulation instability and solitons interactions for a nonlinear Schr\"{o}dinger equation with the sextic operators in an optical fiber. Optical and Quantum Electronics 50 (2018) 340.


\bibitem{gbl-pre-2012} Guo B L, Ling L M, Liu Q P, Nonlinear Schr\"{o}dinger equation: generalized Darboux transformation and rogue wave solutions. Phys. Rev. E 85 (2012) 026607.

\bibitem{hejs-pre-2013} He J S, Zhang H R, Wang L H, Fokas A S, Generating mechanism for higher-order rogue waves. Phys. Rev. E 87 (2013) 052914.



\bibitem{wenxy-chaos-2015} Wen X Y, Yan Z Y, Modulational instability and higher-order rogue waves with parameters modulation in a coupled integrable AB system via the generalized Darboux transformation. Chaos 25 (2015) 123115.


\bibitem{wenxy-jmp-2018} Wen X Y, Yan Z Y, Modulational instability and dynamics of multi-rogue wave solutions for the discrete Ablowitz-Ladik equation. J. Math. Phys. 59 (2018) 073511.


\bibitem{weij-cnsns-2018} Wei J, Wang X, Geng X G, Periodic and rational solutions of the reduced Maxwell-Bloch equations. Commun. Nonlinear Sci. Numer. Simulat. 59 (2018) 1-14.

\bibitem{linglm-pd-2016} Ling L M, Feng B F, Zhu Z N, Multi-soliton, multi-breather and higher-order rogue wave solutions to the complex short pulse equation. Physica D 327 (2016) 13-29.


\bibitem{liuyb-nd-2017} Liu Y B, Mihalache D, He J S, Families of rational solutions of the y-nonlocal Davey-Stewartson II equation. Nonlinear Dyn. 90 (2017) 2445-2455.

\bibitem{chenjc-jmaa-2018} Chen J C, Ma Z Y, Hu Y H, Nonlocal symmetry, Darboux transformation and soliton-cnoidal wave interaction solution for the shallow water wave equation. J. Math. Anal. Appl. 460 (2018) 987-1003.


\bibitem{huangll-cma-2018} Huang L L, Yue Y F, and Chen Y, Localized waves and interaction solutions to a (3+1)-dimensional generalized KP equation. Comput. Math. Appl. 76 (2018) 831-844.

\bibitem{yueyf-aml-2019} Yue Y F, Huang L L, Chen Y, Localized waves and interaction solutions to an extended (3+1)-dimensional Jimbo-Miwa equation. Appl. Math. Lett. 89 (2019) 70-77.

\bibitem{huangll-cnsns-2019} Huang L L, Chen Y, Localized excitations and interactional solutions for the re duce d Maxwell-Bloch equations. Commun. Nonlinear Sci Numer. Simulat. 67 (2019) 237-252.



\bibitem{akhmedievn-1997} Akhmediev N, Ankiewicz A, Solitons, nonlinear pulses and beams. Chapman and Hall, London 1997.

\bibitem{ankiewicza-pre-2010} Ankiewicz A, Soto-Crespo J M, Akhmediev N, Rogue waves and rational solutions of the Hirota equation. Phys. Rev. E 81 (2010) 046602.

\bibitem{ankiewicza-pla-2014} Ankiewicz A, Akhmediev N, Higher-order integrable evolution equation and its soliton solutions. Phys. Lett. A 378 (2014) 358-361.

%
%
%
%
%
%

\bibitem{akhmediev-pla-2011} Akhmediev N, Ankiewicz A, Soto-Crespo J M, Dudley J M, Rogue wave early warning through spectral measurements. Phys. Lett. A 375 (2011) 541-544.

\bibitem{akhmediev-pd-2015} Akhmediev N, Soto-Crespo J M, Devine N, Hoffmann N P, Rogue wave spectra of the Sasa-Satsuma equation. Physica D 294 (2015) 37-42.

\bibitem{bayindir-pre-2016} Bayindir C. Rogue wave spectra of the Kundu-Eckhaus equation. Phys. Rev. E 93 (2016) 062215.

\bibitem{wangx-chaos-2017} Wang X, Liu C, Wang L, Rogue waves and W-shaped solitons in the multiple self-induced transparency system. Chaos 27 (2017) 093106.

\bibitem{wangx-sam-2017} Wang X, Liu C, W-shaped soliton complexes and rogue-wave pattern transitions for the AB system. Superlattices and Microstructures 107 (2017) 299-309.


\bibitem{droquesm-ol-2011} Droques M, Barviau B, Kudlinski A, Taki M, Boucon A, Sylvestre T, Mussot A, Symmetry-breaking dynamics of the modulational instability spectrum. Optics letters 36 (2011) 1359-1361.


\bibitem{sollidr-np-2012} Solli D R, Herink G, Jalali B, Ropers C, Fluctuations and correlations in modulation instability. Nature Photonics 6 (2012) 463-468.

\bibitem{zhaolc-josa-2016} Zhao L C, Ling L M, Quantitative relations between modulational instability and several well-known nonlinear excitations. J. Opt. Soc. Amer. B 33 (2016) 850-856.

\bibitem{liuc-pre-2016} Liu C, Yang Z Y, Zhao L C, Duan L, Yang G Y, Yang W L, Symmetric and asymmetric optical multipeak solitons on a continuous wave background in the femtosecond regime. Phys. Rev. E 94 (2016) 042221.

\bibitem{duanl-pre-2017} Duan L, Zhao L C, Xu W H, Liu C, Yang Z Y, Yang W L, Soliton excitations on a continuous-wave background in the modulational instability regime with fourth-order effects. Phys. Rev. E 95 (2017) 042212.

\bibitem{yangyq-chaos-2015} Yang Y Q, Yan Z Y, Malomed B A, Rogue waves, rational solitons, and modulational instability in an integrable fifth-order nonlinear Schr\"{o}dinger equation. Chaos 25 (2015) 103112.

\bibitem{lip-aml-2018} Li P, Wang L, Kong L Q, Wang X, Xie Z Y, Nonlinear waves in the modulation instability regime for the fifth-order nonlinear Schr\"{o}dinger equation. Appl. Math. Lett. 85 (2018) 110-117.

\end{thebibliography}
\end{document}